\documentstyle[amssymb,prl,aps,epsfig]{revtex}

\begin{document}

\title{Universal properties of cuprate superconductors: Evidence
and implications}
\author{T. Schneider \\
%EndAName
Physik-Institut der Universit\"{a}t Z\"{u}rich, Winterthurerstrasse 190,\\
CH-8057 Z\"{u}rich, Switzerland}

\maketitle

\begin{abstract}

\begin{center}
To be published in:

\bigskip

{\large The Physics of Conventional and Unconventional
Superconductors}

\bigskip

Springer-Verlag, Berlin, ed. K.H. Bennemann and J.B. Ketterson
\end{center}

\end{abstract}

\section{Introduction}

\begin{figure}[tbp]
\centering
\includegraphics[totalheight=5cm]{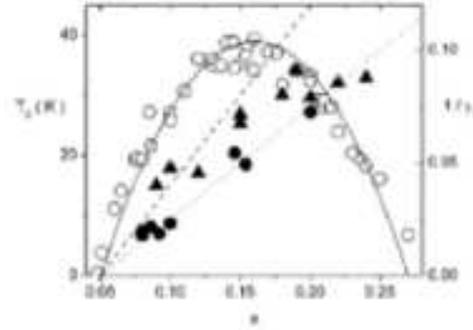}
\caption{Variation of $T_{c}$ (open circles) \protect\cite
{suzuki,nakamura,fukuzumi,willemin,kimura,sasagawa,hoferdis,shibauchi,panagopoulos}
and $\protect\gamma _{T}$ with $x$ for
La$_{2-x}$Sr$_{x}$CuO$_{4}$. Filled circles correspond to
$1/\protect\gamma _{T_{c}}$\ \protect\cite
{suzuki,nakamura,willemin,sasagawa,hoferdis} and filled triangles
to $1/\protect\gamma _{T=0}$
\protect\cite{shibauchi,panagopoulos}. The solid curve is
Eq.(\ref{eq1a}) with $T_{c}^{m}=39 K$. The dashed and dotted lines
follow from Eq.(\ref{eq1b}) with $\protect\gamma _{0,T_{c}}=2$ and
$\protect\gamma _{T_{c}}=1.63$.} \label{fig1}
\end{figure}

Establishing and understanding the phase diagram of cuprate
superconductors in the temperature - dopant concentration plane is
one of the major challenges in condensed matter physics.
Superconductivity is derived from the insulating and
antiferromagnetic parent compounds by partial substitution of ions
or by adding or removing oxygen. For instance La$_{2}$CuO$_{4}$
can be doped either by alkaline earth ions or oxygen to exhibit
superconductivity. The empirical phase diagram of
La$_{2-x}$Sr$_{x}$CuO$_{4}$ \cite
{suzuki,nakamura,fukuzumi,willemin,kimura,sasagawa,hoferdis,shibauchi,panagopoulos}
depicted in Fig.\ref{fig1} shows that after passing the so called
underdoped limit $\left( x_{u}\approx 0.047\right) $, $T_{c}$
reaches its maximum value $T_{c}^{m}$ at $x_{m}\approx 0.16$. With
further increase of $x$, $T_{c}$ decreases and finally vanishes in
the overdoped limit $x_{o}\approx 0.273$. This phase transition
line is thought to be a generic property of cuprate
superconductors \cite{tallon} \ and is well described by the
empirical relation
\begin{equation}
T_{c}\left( x\right) =T_{c}\left( x_{m}\right) \left( 1-2\left(
\frac{x}{x_{m}}-1\right) ^{2}\right) =\frac{2T_{c}\left(
x_{m}\right) }{x_{m}^{2}} \left( x-x_{u}\right) \left(
x_{o}-x\right) ,\ \ x_{m}=0.16, \label{eq1a}
\end{equation}
proposed by Presland {\em et al}.\cite{presland}. Approaching the
endpoints along the axis $x$, La$_{2-x}$Sr$_{x}$CuO$_{4}$
undergoes at zero temperature doping tuned quantum phase
transitions. As their nature is concerned, resistivity
measurements reveal a quantum superconductor to insulator (QSI)
transition in the underdoped limit\cite
{fukuzumi,polen,book,klosters,momono} and in the overdoped limit a
quantum superconductor to normal state (QSN)
transition\cite{momono}.

\begin{figure}[tbp]
\centering
\includegraphics[totalheight=5cm]{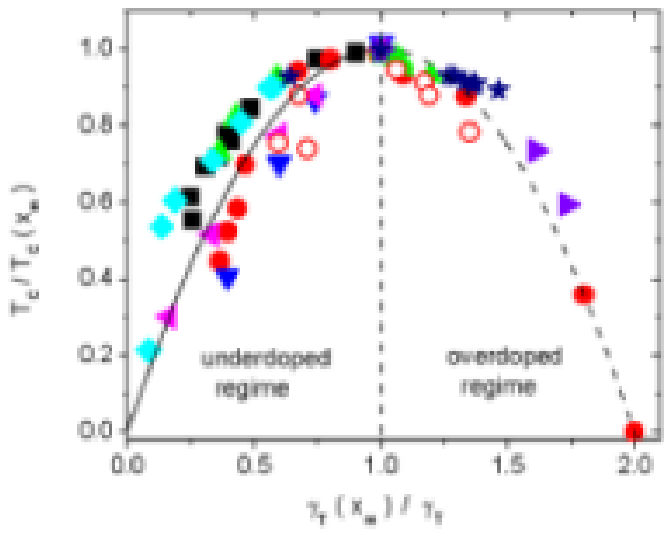}
\caption{$T_{c}/T_{c}\left( x_{m}\right) $ versus $\protect\gamma
_{T}\left(
x_{m}\right) /$ $\protect\gamma _{T}$ for La$_{2-x}$Sr$_{x}$CuO$_{4}$ ($%
\bullet $, $T_{c}\left( x_{m}\right) =37 K$, $\protect\gamma
_{T_{c}}\left(
x_{m}\right) =20$) \protect\cite{suzuki,nakamura,willemin,sasagawa,hoferdis}%
\ , ($\bigcirc $, $T_{c}\left( x_{m}\right) =37 K$, $\protect\gamma %
_{T=0}\left( x_{m}\right) =14.9$) \protect\cite{shibauchi,panagopoulos}, HgBa%
$_{2}$CuO$_{4+\protect\delta }$ ($\blacktriangle $, $T_{c}\left(
x_{m}\right) =95.6 K$, $\protect\gamma _{T_{c}}\left( x_{m}\right)
=27$)
\protect\cite{hoferhg}, Bi$_{2}$Sr$_{2}$CaCu$_{2}$O$_{8+\protect\delta }$ ($%
\bigstar $, $T_{c}\left( x_{m}\right) =84.2 K$, $\protect\gamma %
_{T_{c}}\left( x_{m}\right) =133$)\protect\cite{watauchi}, YBa$_{2}$Cu$_{3}$O%
$_{7-\protect\delta }$ ($\blacklozenge $, $T_{c}\left(
x_{m}\right) =92.9 K$,
$\protect\gamma _{T_{c}}\left( x_{m}\right) =8$) \protect\cite{chien123}, YBa%
$_{2}$(Cu$_{1-y}$Fe$_{y}$)$_{3}$O$_{7-\protect\delta }$ ($\blacksquare $, $%
T_{c}\left( x_{m}\right) =92.5 K$, $\protect\gamma _{T_{c}}\left(
x_{m}\right) =9$)\protect\cite{chienfe}, Y$_{1-y}$Pr$_{y}$Ba$_{2}$Cu$_{3}$O$%
_{7-\protect\delta }$ ($\blacktriangledown $, $T_{c}\left( x_{m}\right) =91 K$%
, $\protect\gamma _{T_{c}}\left( x_{m}\right)
=9.3$)\protect\cite{chienpr},
BiSr$_{2}$Ca$_{1-y}$Pr$_{y}$Cu$_{2}$O$_{8}$ ($\blacktriangleleft $, $%
T_{c}\left( x_{m}\right) =85.4 K$, $\protect\gamma _{T=0}\left(
x_{m}\right)
=94.3$) \protect\cite{sun} and YBa$_{2}$(Cu$_{1-y}$ Zn$_{y}$)$_{3}$O$_{7-%
\protect\delta }$ ($\blacktriangleright $, $T_{c}\left( x_{m}\right) =92.5 K$%
, $\protect\gamma _{T=0}\left( x_{m}\right) =9$)
\protect\cite{panagopzn}. The solid and dashed curves are
Eq.(\ref{eq4}), marking the flow from the maximum $T_{c}$ to QSI
and QSN criticality, respectively. }
\label{fig2}
\end{figure}

Another essential experimental fact is the doping dependence of
the anisotropy. In tetragonal cuprates it is defined as the ratio
$\gamma =\xi _{ab}/\xi _{c}$ of the correlation lengths parallel
$\left( \xi _{ab}\right) $ and perpendicular $\left( \xi
_{c}\right) $ to CuO$_{2}$ layers (ab-planes). In the
superconducting state it can also be expressed as the ratio
$\gamma =\lambda _{c}/\lambda _{ab}$ of the London penetration
depths due to supercurrents flowing perpendicular ($\lambda _{c}$
) and parallel ($ \lambda _{ab}$ ) to the ab-planes. Approaching a
nonsuperconductor to superconductor transition $\xi $ diverges,
while in a superconductor to nonsuperconductor transition $\lambda
$ tends to infinity. In both cases, however, $\gamma $ remains
finite as long as the system exhibits anisotropic but genuine 3D
behavior. There are two limiting cases: $\gamma =1$ characterizes
isotropic 3D- and $\gamma =\infty $ 2D-critical behavior. An
instructive model where $\gamma $ can be varied continuously is
the anisotropic 2D Ising model\cite{onsager}. When the coupling in
the y direction goes to zero, $\gamma =\xi _{x}/\xi _{y}$ becomes
infinite, the model reduces to the 1D case and $T_{c}$ vanishes.
In the Ginzburg-Landau description of layered superconductors the
anisotropy is related to the interlayer coupling. The weaker this
coupling is, the larger $\gamma $ is. The limit $\gamma =\infty $
is attained when the bulk superconductor corresponds to a stack of
independent slabs of thickness $d_{s}$. With respect to
experimental work, a considerable amount of data is available on
the chemical composition dependence of $\gamma $ . At $T_{c}$ it
can be inferred from resistivity ($\gamma =\xi _{ab}/\xi
_{c}=\sqrt{\rho _{ab}/\rho _{c}}$) and magnetic torque
measurements, while in the superconducting state it follows from
magnetic torque and penetration depth ($\gamma =\lambda
_{c}/\lambda _{ab}$) data. In Fig.\ref{fig1} we included the
doping dependence of $1/\gamma _{T}$ evaluated at $T_{c}$ ($\gamma
_{T_{c}}$) and $T=0$ ($\gamma _{T=0}$). As the dopant
concentration is reduced, $\gamma _{T_{c}}$ and $\gamma _{T=0}$
increase systematically, and tend to diverge in the underdoped
limit. Thus the temperature range where superconductivity occurs
shrinks in the underdoped regime with increasing anisotropy. This
competition between anisotropy and superconductivity raises
serious doubts whether 2D mechanisms and models, corresponding to
the limit $\gamma _{T}=\infty $, can explain the essential
observations of superconductivity in the cuprates. From
Fig.\ref{fig1} it is also seen that $\gamma _{T}\left( x\right) $
is well described by
\begin{equation}
\gamma _{T}\left( x\right) =\frac{\gamma _{T,0}}{x-x_{u}}.
\label{eq1b}
\end{equation}
Having also other cuprate families in mind, it is convenient to
express the dopant concentration in terms of $T_{c}$. From
Eqs.(\ref{eq1a}) and(\ref {eq1b}) we obtain the correlation
between $T_{c}$ and $\gamma _{T}$:
\begin{equation}
\frac{T_{c}}{T_{c}\left( x_{m}\right) }=1-\left( \frac{\gamma
_{T}\left( x_{m}\right) }{\gamma _{T}}-1\right) ^{2},\ \ \gamma
_{T}\left( x_{m}\right) =\frac{\gamma _{T,0}}{x_{m}-x_{u}}
\label{eq1c}
\end{equation}
Provided that this empirical correlation is not merely an artefact
of La$_{2-x}$Sr$_{x}$CuO$_{4}$, it gives a universal perspective
on the interplay of anisotropy and superconductivity, among the
families of cuprates, characterized by $T_{c}\left( x_{m}\right) $
and $\gamma _{T}\left( x_{m}\right) $. For this reason it is
essential to explore its generic validity. In practice, however,
there are only a few additional compounds, including
HgBa$_{2}$CuO$_{4+\delta }$\cite{hoferhg} and
Bi$_{2}$Sr$_{2}$CuO$_{6+\delta }$, for which the dopant
concentration can be varied continuously throughout the entire
doping range. It is well established, however, that the
substitution of magnetic and nonmagnetic impurities, depress
$T_{c}$ of cuprate superconductors very
effectively\cite{xiao,tarascon}. To compare the doping and
substitution driven variations of the anisotropy, we depicted in
Fig.\ref{fig2} the plot $T_{c}/T_{c}\left( x_{m}\right) $ versus
$\gamma _{T}\left( x_{m}\right) /$ $\gamma _{T}$ for a variety of
cuprate families. The collapse of the data on the parabola, which
is the empirical relation (\ref{eq1c}), reveals that this scaling
form appears to be universal. Thus, given a family of cuprate
superconductors, characterized by $T_{c}\left( x_{m}\right) $ and
$\gamma _{T}\left( x_{m}\right) $, it gives a universal
perspective on the interplay between anisotropy and
superconductivity.

\begin{figure}[tbp]
\centering
\includegraphics[totalheight=5cm]{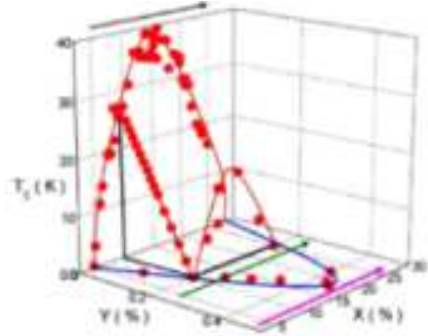}
\caption{Phase diagram of \
La$_{2-x}$Sr$_{x}$Cu$_{1-y}$Zn$_{y}$O$_{4}$ . The blue solid curve
corresponds to $y_{c}\left( x\right) $, a line of quantum phase
transitions. The pink arrow marks the doping tuned insulator to
metal crossover and the green arrow marks a path where a QSI and
QSN transition occurs. Experimental data taken from Momono {\em et
al.} \protect\cite{momono}.} \label{fig3}
\end{figure}

\begin{figure}[tbp]
\centering
\includegraphics[totalheight=5cm]{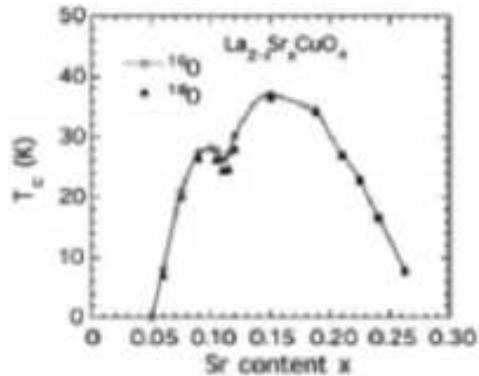}
\caption{$T_{c}$ ($^{16}$O ) and $T_{c}$ ($^{18}$O ) versus $x$
for La$_{2-x} $Sr$_{x}$CuO$_{4}$ . From Guo-Meng Zhao {\em et
al.}\protect\cite{zhao214iso}. }
\label{fig4}
\end{figure}

The effect of a substitution for Cu by other magnetic or
nonmagnetic metals was also investigated extensively
\cite{momono,xiao,tarascon}. A result common to all of these
studies is that $T_{c}$ is suppressed in the same manner,
independent of wether the substituent is magnetic or nonmagnetic.
For this reason, the phase diagram of \
La$_{2-x}$Sr$_{x}$Cu$_{1-y}$Zn$_{y}$O$_{4}$, depicted in
Fig.\ref{fig3}, applies quite generally. Apparently, the
substituent axis (y) extends the complexity and richness of the
phase diagram considerably. The blue curve corresponds to a line
of quantum phase transitions, given by $y_{c}\left( x\right) $.
The pink arrow marks the doping tuned insulator to metal crossover
and the green arrow corresponds to a path along which a QSI and
QSN transition occurs. From Fig. \ref{fig4} it can be inferred
that isotope substitution, though much less effective, has
essentially the same effect. $T_{c}$ is lowered and the underdoped
limit $x_{u}$ shifts to some $y_{c}\left( x\right) $. This
suggests that substitution induced local distortions, rather than
magnetism, is the important factor. For $y>y_{c}\left( x\right) $
superconductivity is suppressed due to the destruction of phase
coherence.

\begin{figure}[tbp]
\centering
\includegraphics[totalheight=5cm]{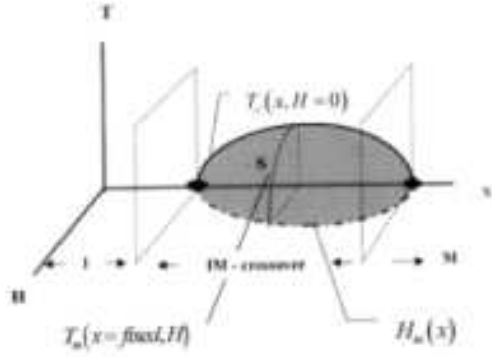}
\caption{Schematic ($x,H,T$)-phase diagram. There is the
superconducting phase (S), bounded by the zero-field transition
line, $T_{c}\left( x,H=0\right) $, the critical lines of \ the
vortex melting or vortex glass to vortex fluid transitions, \
$T_{m}\left( x=\text{fixed},H\right) $ and the line of quantum
critical points, $H_{m}\left( x,T=0\right) $. Along this line
superconductivity is suppressed and the critical endpoints
coincide with the 2D-QSI- and 3D-QSN- critical points at $x_{u}$
and $x_{o}$, respectively.} \label{fig5}
\end{figure}

The point of reference for magnetic field tuned transitions is
embodied in the schematic phase diagram \ shown in Fig.\ref{fig5}.
It is strongly affected by the combined effect of pinning, thermal
and quantum fluctuations, anisotropy and dimensionality
\cite{blatter}. In clean cuprates and close to $T_{c}\left(
x,H=0\right) $ thermal fluctuations are thought to be responsible
for the existence of a first-order vortex melting transition. In
the presence of disorder, however, the long-range order of the
vortex lattice is destroyed and the vortex solid becomes a glass
\cite {vortexglass}. Since a sufficiently large magnetic field
suppresses superconductivity, due to the destruction of phase
coherence, there is a line $H_{m}\left( x\right) $ of quantum
phase transitions, connecting the zero field QSI and QSN
transitions. Indeed, recent experiments revealed that
sufficiently-intense magnetic fields suppress superconductivity
and mediate a metal to insulator (MI) crossover
\cite{ando,boebinger,segawa,ono}.

\begin{figure}[tbp]
\centering
\includegraphics[totalheight=5cm]{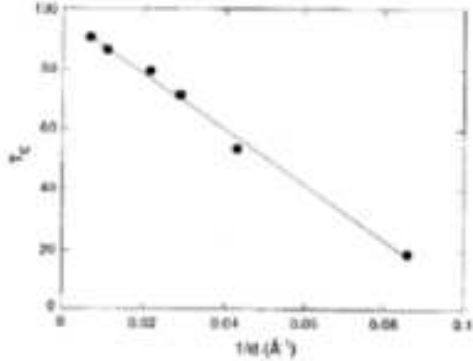}
\caption{Zero resistance $T_{c}$ versus $1/d$ \ of
YBa$_{2}$Cu$_{3}$O$_{7-\protect\delta \text{ }}$ layers of
thickness $d$ separated by 16 PrBa$_{2}$Cu$_{3}$O$_{7\text{
}}$unit cells. Taken from Goodrich {\em et al.}
\protect\cite{goodrich}. The straight line is a linear fit to
Eq.(\ref{eq1d}). } \label{fig6}
\end{figure}

\begin{figure}[tbp]
\centering
\includegraphics[totalheight=6cm]{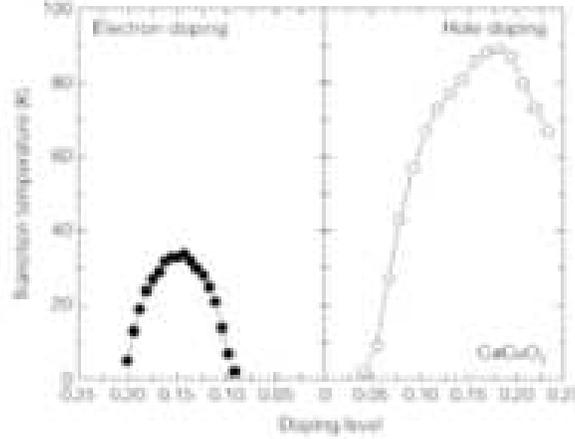}
\caption{($T_{c}$, $x$) phase diagram of CaCuO$_{2}$. A maximum
$T_{c}$ of $89$ and $34$ K is obtained at optimal hole and
electron doping, respectively. Doping level x is defined as number
of charge carriers per CuO$_{2}$ unit. Taken from Sch\"{o}n {\em
et al.}\protect\cite{schon2}.}
\label{fig7}
\end{figure}

The QSI transition can also be traversed in films by changing
their thickness\cite{book}. An instructive example are the
measurements on YBa$_{2}$Cu$_{3}$O$_{7-\delta }$ slabs of
thickness $d$ separated by 16 unit cells ($\approx 187\ A$) of
PrBa$_{2}$Cu$_{3}$O$_{7}$. Due to their large separation the
YBa$_{2}$Cu$_{3}$O$_{7-\delta }$ slabs are essentially uncoupled.
As shown in Fig.\ref{fig6}, $T_{c}$ was found to vary with the
thickness $d$ of the YBa$_{2}$Cu$_{3}$O$_{7-\delta }$ slabs as
\begin{equation}
\frac{T_{c}\left( d\right) }{T_{c}^{bulk}}=\frac{d_{s}}{d}\left(
\frac{d}{d_{s}}-1\right), \label{eq1d}
\end{equation}
with $T_{c}^{bulk}=91\ K$ and $d_{s}=10.1\ A$\cite{goodrich}.
$d_{s}$ is the critical film thickness below which
superconductivity is lost. Although the decrease of $T_{c}$ is
partially due to the 3D-2D crossover, the occurrence of the QSI
transition points to the dominant role of disorder and quantum
fluctuations. For this reason it is conceivable that in
sufficiently clean films superconductivity may also occur at and
slightly below $d_{s}$. Recently this has been achieved by
inducing charges by the field-effect technique, whereby the mobile
carrier density can be varied continuously and reversibly, without
introducing additional sources of disorder\cite
{ahn,kawaharo,schon2}. By establishing a voltage difference
between a metallic electrode and the crystal, charge can be added
or removed from the CuO$_{2}$ layers: positive voltage injects
electrons, and a negative voltage injects holes. This allows to
vary both electron and hole charge densities over a considerable
range of interest. Using this technique Sch\"{o}n {\em et
al.}\cite{schon2} converted an insulating thin slab of \
CaCuO$_{2}$ into a superconductor. The success of this technique
relies on the high quality films and on the quality of the
interfaces between the film material and the metallic electrode.
The ($T_{c}$, $x$) phase diagram for CaCuO$_{2}$ is shown
Fig.\ref{fig7}. The doping level was calculated from the
independently measured capacitance of the gate dielectric,
assuming that all the charge is located in a single CuO$_{2}$
layer. The diagram exhibits apparent electron and hole
superconductivity, and bears a strong resemblance to results
observed in other cuprates, in qualitative agreement with the
empirical correlation (\ref{eq1c}) for bulk systems. Note that the
finite temperature superconductor to normal state transition falls
into the 2D-XY universality class. Thus $T_{c}$ is the
Kosterlitz-Thouless transition temperature. At low doping levels
($x<0.1$) a logarithmic divergence of the resistivity was found.
This observation for n- as well as p-type doping of CaCuO$_{2}$
are in accordance with a 2D-QSI transition in the underdoped
limit. Although superconductivity has been observed previously in
SrCuO$_{2}$/BaCuO$_{2}$\cite{norton} and in
BaCuO$_{2}$/CaCuO$_{2}$\cite{balestrino} superlattices with a
maximum $T_{c}$ value of about 80 K, the field effect doping is
not limited by dopant solubility and compositional stability. This
technique allows to study the properties of a given material as a
function of the dopant concentration without introducing
additional disorder and additional defects. Because the charge
carriers appear to be confined in a very thin layer, what's seen
is superconductivity in 2D. Thus, due to the reduced
dimensionality fluctuations will be enhanced over the full range
of dopant concentrations. The phase diagram displayed in
Fig.\ref{fig7} also points to a 2D-QSN transition in the overdoped
limit. In any case, this phase diagram clearly reveals the
survival of superconductivity in field effect doped CaCuO$_{2}$ in
the 2D-limit, while it disappears in the bulk (see e.g.
Fig.\ref{fig2}) with increasing anisotropy and in chemically doped
films at a critical thickness $d_{s}$, presumably controlled by
disorder.

This review aims to analyze the empirical correlations and phase
diagrams from the point of view of thermal and quantum critical
phenomena, to identify the universal properties, the effective
dimensionality and the associated crossover phenomena. In view of
the mounting evidence for 3D-XY-universality close to optimum
doping \cite
{hoferdis,book,ariosa,tskel,hubbard,kamal,pasler,tseuro,houston},
we concentrate here on the thermodynamic and ground state
properties emerging from the QSI- and QSN- transitions, including
the associated crossover phenomena. For this purpose we invoke the
scaling theory of quantum critical phenomena\cite{book,kim}.

Zero temperature phase transitions in quantum systems differ
fundamentally from their finite temperature counterparts in that
their thermodynamics and dynamics are inextricably mixed.
Nevertheless, by means of the path integral formulation of quantum
mechanics, one can view the statistical mechanics of D-dimensional
$T=0$ quantum system as the statistical mechanics of a $D+z$
dimensional classical system with a fake temperature which is some
measure of the dynamics, characterized by the dynamic critical
exponent $z$. This allows one to apply the scaling theory
developed for classical critical phenomena to quantum criticality.
In particular this leads to an understanding of the low $T$ and
crossover behavior close to quantum phase transitions and to
universal relations between various properties. Evidence for power
law behavior should properly consist of data that cover several
decades in the parameters to provide reliable estimates for the
critical exponents. In cuprate superconductors, the various power
laws span at best one decade. Accordingly, more extended
experimental data are needed to determine the critical exponents
of the quantum phase transitions. Nevertheless, irrespective of
their precise value, the evidence for scaling and with that for
data collapse exists. It uncovers the relationship between various
properties and the significance of the empirical correlations and
offers an understanding of the doping, substitution and magnetic
field tuned quantum phase transition points and lines (see
Figs.\ref{fig1}, \ref{fig3} and \ref{fig5}). Evidently, the
anisotropy, the associated dimensional crossover and the scaling
relations between various properties close to the OSI and QSN
criticality provide essential constraints for the understanding of
the phase diagrams and the microscopic theory of superconductivity
in these materials.

Note that this scenario is not incompatible with the zoo of
microscopic models, relying on competing order parameters \cite
{sachdev1,monthoux,castellani,varma,zhang,kivelson,randeria,capara,chakravarty,voijta,anderson}.
Here it is assumed that in the doping regime where
superconductivity occurs, competing fluctuations, including
antiferromagnetic and charge fluctuations, can be integrated out.
The free-energy density is then a functional of a complex scalar,
the order parameter of the superconducting phase, only. Given the
generic phase diagrams (Figs.\ref{fig1}, \ref{fig3} and
\ref{fig5}) the scaling theory of finite temperature and quantum
critical phenomena leads to predictions, including the universal
properties, which can be confronted with experiment. As it stands,
the available experimental data appears to be fully consistent
with a single complex scalar order parameter, a doping tuned
dimensional crossover and a doping, substitution or magnetic field
driven suppression of superconductivity, due to the loss of phase
coherence. When the evidence for this scenario persists,
antiferromagnetic and charge fluctuations turn out to be
irrelevant close to criticality. Moreover, it implies that a
finite transition temperature and superfluid aerial superfluid
density in the ground state, require in chemically doped systems a
finite anisotropy. The important conclusion there is that a finite
superfluid density in the ground state of bulk cuprates oxides is
unalterably linked to an anisotropic but 3D condensation
mechanism. Thus despite the strongly two-dimensional layered
structure of cuprate superconductors, a finite anisotropy
associated with the third dimension, perpendicular to the
CuO$_{2}$ planes, is an essential factor in mediating
superfluidity.

The paper is organized as follows. Sec.II is devoted to the finite
temperature critical behavior. Since a substantial review on this
topic is available\cite{book}, we concentrate on the specific
heat. In Sec.II.A we sketch the scaling theory of finite
temperature critical phenomena in anisotropic superconductors
falling into the 3D-XY universality class. This leads naturally to
universal critical amplitude combinations, involving the
transition temperature and the critical amplitudes of specific
heat, correlation lengths and penetration depths. The universality
class to which the cuprates belong is thus not only characterized
by its critical exponents but also by various critical-point
amplitude combinations which are equally important. Indeed, though
these amplitudes depend on the dopant concentration, substitution
etc., their universal combinations do not. Evidence for 3D-XY
universality and their implication for the vortex melting
transition is presented in Sec.II.B. Here we also discuss the
limitations arising from the inhomogeneities and the anisotropy,
which render it difficult to observe 3D-XY critical behavior along
the entire phase transition line $T_{c}\left( x\right) $
(Fig.\ref{fig1}) or on the entire surface $T_{c}\left( x,y\right)
$ (Fig.\ref{fig3}).

In Sec.III we examine the quantum phase transitions and the
associated crossover phenomena. The scaling theory of quantum
phase transitions\cite {kim}, extended to anisotropic
superconductors\cite{book}, is reviewed in Sec.III.A. Essential
predictions include a universal amplitude relation in D=2
involving the transition temperature and the zero temperature
in-plane penetration depth, as well as a fixed value of the
in-plane sheet conductivity. Moreover we explore the scaling
properties of transition temperature, penetration depths,
correlation lengths, anisotropy and specific heat coefficient at
2D-QSI and 3D-QSN criticality. In Sec.III.B we confront these
predictions with the empirical correlations (\ref{eq1a}),
(\ref{eq1b}) and (\ref{eq1c}) and pertinent experiments. Although
the experimental data are rather sparse, in particular close to
the 3D-QSN transition, we observe a flow pattern pointing
consistently to a 2D-QSI transition with $z=1$ and $\overline{\nu
}\approx 1$, and 3D-QSN criticality with $z=2$ and $\overline{\nu
}\approx 1/2$. $z$ is the dynamic critical exponent and
$\overline{\nu }$ the correlation length exponent. The estimates
for the 2D-QSI transition coincide with the theoretical prediction
for a 2D disordered bosonic system with long-range Coulomb
interactions\cite {fisher,herbut}. This reveals that in cuprate
superconductors the loss of phase coherence, due to the
localization of Cooper pairs, is responsible for the 2D-QSI
transition. On the other hand, $z=2$ and $\overline{\nu }\approx
1/2$ point to a 3D-QSN critical point, compatible with a
disordered metal to d-wave superconductor transition at weak
coupling\cite{herbutd}. Here the disorder destroys
superconductivity, while at the 2D-QSI transition it destroys
superfluidity. A characteristic feature of the 2D-QSI transition
is its rather wide and experimentally accessible critical region.
For this reason we observe consistent evidence that it falls into
the same universality class as the onset of superfluidity in
$^{4}$He films in disordered media, corrected for the
long-rangeness of the Coulomb interaction. As also discussed in
this section, the existence of 2D-QSI and 3D-QSN critical points
implies a doping and substitution tuned dimensional crossover. A
glance to Fig.\ref{fig2} shows that it is due to the dependence of
the anisotropy on doping and substitution. An important
implication is, that despite the small fraction, which the third
dimension contributes to the superfluid energy density in the
ground state, a finite transition temperature and superfluid
density in bulk cuprates is unalterably linked to a finite
anisotropy. Thus, despite their strongly two-dimensional layered
structure, a finite anisotropy associated with the third
dimension, perpendicular to the CuO$_{2}$ planes, is an essential
factor in mediating pair condensation. This points unambiguously
to the conclusion that theories formulated for a single CuO$_{2}$
plane cannot be the whole story. Moreover, the evidence for the
flow to 2D-QSI criticality also implies that the standard
Hamiltonian for layered superconductors\cite{glazman} is
incomplete. Although its critical properties fall into the 3D-XY
universality class, disorder and quantum fluctuations must be
included to account for the flow to 2D-QSI and 3D-QSN criticality.

Sec.IV is devoted to the magnetic field tuned quantum phase
transitions. Contrary to finite temperature, disorder is an
essential ingredient at $T=0$. It destroys superconductivity at
3D-QSN criticality and superfluidity at 2D-QSI critical points. On
the other hand, superconductivity is also destroyed by a
sufficiently large magnetic field. Accordingly, one expects a line
$H_{m}\left( x\right) $ of quantum phase transitions, connecting
the zero field 2D-QSI and 3D-QSN transitions (see Fig.\ref{fig5}).
The relevance of disorder at this critical endpoints suggests a
line of quantum vortex glass to vortex fluid transitions. Although
the available experimental data is rather sparse, it points to the
existence of a quantum critical line $H_{m}\left( x\right) $ and
2D localization, consistent with 2D-QSI criticality.

In Sec.V we treat cuprates with reduced dimensionality.
Empirically it is well established that a quantum superconductor
to insulator transition in thin films can also be traversed by
reducing the film thickness. There is a critical film thickness
($d_{s}$) where $T_{c}$ vanishes and below which disorder destroys
superconductivity\cite{book}. In chemically doped cuprates the
critical thickness is comparable to the c-axis lattice constant.
Moreover, the empirical correlation (\ref{eq1c}), displayed in
Fig.\ref{fig2}, implies that in the bulk superconductivity
disappears in the 2D limit. Thus, the combined effect of disorder
and quantum fluctuations appears to prevent the occurrence of
strictly 2D superconductivity. For this reason it is conceivable
that in sufficiently clean films superconductivity may also occur
at and below this value of $d_{s}$. As aforementioned, this has
been achieved by field effect doped CaCuO$_{2}$\cite {schon2}.
Although more detailed studies are needed to uncover the
characteristic 2D critical properties in field effect doped
cuprates, the phase diagram of CaCuO$_{2}$ clearly points to a
2D-QSI transition in the underdoped and a 2D-QSN transition in the
overdoped limit. It also implies that in field effect doped
cuprates superconductivity survives in the 2D limit, while in the
bulk and chemically doped films it disappears. Since chemically
doped materials with different carrier densities also have varying
amounts of disorder, the third dimension appears to be needed to
delocalize the carriers and to mediate superfluidity. The
comparison with other layered superconductors, including organics
and dichalcogenides is made in Sec.VI.

\section{Finite temperature critical behavior}

\subsection{Sketch of the scaling predictions}

In superconductors the order parameter is a complex scaler, but it
can also be viewed as a two component vector (XY). Supposing that
sufficiently close to the phase transition line, separating the
superconducting and non superconducting phase, 3D-XY-fluctuations
dominate, the scaling form of the singular part of the bulk free
energy density adopts then the form \cite {book,hohenberg} \
\begin{equation}
f_{s}=-k_{B}TQ_{3}^{\pm }\left( \xi _{x}^{\pm }\xi _{y}^{\pm }\xi
_{z}^{\pm }\right) ^{-1}  \label{eq1}
\end{equation}
where $\xi _{i}^{\pm }$ is the correlation length diverging as
\begin{equation}
\xi _{i}^{\pm }=\xi _{i,0}^{\pm }\left| t\right| ^{-\nu },\
i=x,y,z,\ \pm \ =sign\left( t\right) ,\ \ t=\frac{T-T_{c}}{T_{c}}
\label{eq2}
\end{equation}
and $Q_{3}^{\pm }$ are universal constants. In this context it
should be kept in mind, that in superconductors the pairs carry a
non zero charge in addition to their mass and the charge $\left(
\Phi _{0}=hc/2e\right) $ couples the order parameter to the
electromagnetic field via the gradient term in the Ginzburg-Landau
Hamiltonian. In extreme type II superconductors, however, the
coupling to vector potential fluctuations appears to be
weak\cite{dsfisher}, but nonetheless, in principle, these
fluctuations drive the system very close to criticality, to a
charged critical point\cite{herbuttes}. In any case,
inhomogeneities in cuprate superconductors appear to prevent from
entering this regime, due to the associated finite size
effect\cite{book}. For these reasons, the neglect of vector
potential fluctuations appears to be justified and the critical
properties at finite temperature are then those of the 3D-XY -
model, reminiscent to the lamda transition in superfluid helium,
extended to take the anisotropy into account\cite{book,ariosa}.

In the superconducting phase the order parameter $\Psi $ can be
decomposed into a longitudinal ($\Psi _{0}+\Psi _{lo}$) and
transverse ($\Psi _{tr}$) part:
\begin{equation}
\Psi =\Psi _{0}+\Psi _{lo}+i\Psi _{tr},  \label{eq2a}
\end{equation}
where $\Psi _{0}=\left\langle \Psi \right\rangle $ is chosen to be
real. At long wavelength and in the superconducting phase the
transverse fluctuations dominate and the correlations do not decay
exponentially, but according to a power law\cite{book,hohenberg}.
This results in an inapplicability of the usual definitions of a
correlation length below $T_{c}$. However, in terms of the
helicity modulus, which is a measure of the response of the system
to a phase-twisting field, a phase coherence length can be
defined\cite {helfisher}. This length diverges at critical points
and plays the role of the standard correlation length below
$T_{c}$. In the presence of a phase twist of wavenumber $k_{i}$,
the singular part of the free energy density adopts the scaling
form
\begin{equation}
f_{s}=-\frac{k_{B}TQ_{3}^{-}}{\xi _{x}^{tr}\xi _{y}^{tr}\xi
_{z}^{tr}}\Phi \left( k_{x}\sqrt{\xi _{y}^{tr}\xi
_{z}^{tr}},k_{y}\sqrt{\xi _{x}^{tr}\xi _{z}^{tr}},k_{z}\sqrt{\xi
_{x}^{tr}\xi _{y}^{tr}}\right) ,  \label{eq2b}
\end{equation}
yielding for the helicity modulus the expression
\begin{equation}
\Upsilon _{i}=-=\left. \frac{\partial ^{2}f_{s}}{\partial
k_{i}^{2}}\right| _{{\bf k}={\bf 0}}=\frac{k_{B}TQ_{3}^{-}}{\xi
_{i}^{tr}}\left. \frac{\partial ^{2}\Phi }{\partial
k_{i}^{2}}\right| _{{\bf k}={\bf 0}} \label{eq2c}
\end{equation}
where the normalization, $Q_{3}^{-}\left( \partial ^{2}\Phi
/\partial k_{i}^{2}\right) _{{\bf k}={\bf 0}}$ $=1$, has been
chosen. At $T_{c}$ this leads to the universal relation
\begin{equation}
\left( k_{B}T_{c}\right) ^{3}=\xi _{x0}^{tr}\xi _{y0}^{tr}\xi
_{z0}^{tr}\Upsilon _{x0}\Upsilon _{y0}\Upsilon _{z0}=\left(
\frac{\Phi _{0}^{2}}{16\pi ^{3}}\right) ^{3}\frac{\xi
_{x0}^{tr}\xi _{y0}^{tr}\xi _{z0}^{tr}}{\lambda _{x0}^{2}\lambda
_{y0}^{2}\lambda _{z0}^{2}} \label{eq2d}
\end{equation}
and the definition of the phase coherence lengths, also referred
to as the transverse correlation lengths. The critical amplitudes
of the transverse correlation length, $\xi _{i0}^{tr}$, helicity
modulus, $\Upsilon _{i0}$ and penetration depth, $\lambda _{x0}$,
are the defined as
\begin{equation}
\xi _{i}^{tr}=\xi _{i0}^{tr}\left| t\right| ^{-\nu }\text{,
}\Upsilon _{i}=\Upsilon _{i0}\left| t\right| ^{\nu },\ \ \lambda
_{i}=\lambda _{i0}\left| t\right| ^{-\nu /2}.  \label{eq2e}
\end{equation}
\ The relationship between helicity modulus and penetration depth,
used in Eq.(\ref{eq2d}), is obtained as follows. From the
definition of the supercurrent
\begin{equation}
j_{i}=c\frac{\delta f_{s}}{\delta A_{i}},  \label{eq2f}
\end{equation}
where ${\bf A}$ is the vector potential and $c$ the speed of
light, we obtain for the magnetic penetration depth the expression
\begin{equation}
\frac{1}{\lambda _{i}^{2}}=-\left( \frac{4\pi
j_{i}}{cA_{i}}\right) _{{\bf A=0}}=-\frac{16\pi ^{3}}{\Phi
_{0}^{2}}\left. \frac{\partial ^{2}f_{s}}{\partial k^{2}}\right|
_{k=0}=\frac{16\pi ^{3}}{\Phi _{0}^{2}}\Upsilon _{i}, \label{eq2g}
\end{equation}
by imposing the twist, $k_{i}=2\pi A_{i}/\Phi _{0}$.

Noting then that the transverse correlation function decays
algebraically,
\begin{equation}
S_{tr}\left( R_{i}\right) =\left\langle \Psi _{tr}\left(
R_{i}\right) \Psi _{tr}\left( 0\right) \right\rangle \propto
\left\langle \Psi
_{0}\right\rangle ^{2}\frac{\sqrt{\xi _{j}^{tr}\xi _{j^{\prime }}^{tr}}}{%
R_{i}},\ j\neq j^{\prime }\neq i,  \label{eq2h}
\end{equation}
it is readily seen that the length scales $\xi ^{-}$ correspond to
the real space counterparts of the transverse correlation length
defined in terms of the helicity modulus. These length scales are
related by
\begin{equation}
\xi _{i}^{-}=\sqrt{\xi _{j}^{tr}\xi _{j^{\prime }}^{tr}},\ j\neq
j^{\prime }\neq i  \label{eq2i}
\end{equation}
so that,
\begin{equation}
\xi _{x0}^{-}\xi _{y0}^{-}\xi _{z0}^{-}=\xi _{x0}^{tr}\xi
_{y0}^{tr}\xi _{z0}^{tr}.  \label{eq2k}
\end{equation}

From the singular behavior of the specific heat
\begin{equation}
\frac{C}{Vk_{B}}=-\frac{T}{k_{B}}\frac{\partial ^{2}f_{s}}{\partial t^{2}}%
\approx \frac{A^{\pm }}{\alpha }\left| t\right| ^{-\alpha },
\label{eq3}
\end{equation}
it the follows that the combination of critical amplitudes
\begin{equation}
\left( R^{\pm }\right) ^{3}=A^{\pm }\xi _{x}^{\pm }\xi _{y}^{\pm
}\xi _{z}^{\pm }=A^{\pm }\xi _{x0}^{tr}\xi _{y0}^{tr}\xi
_{z0}^{tr}=-Q_{3}^{\pm }\alpha \left( 1-\alpha \right) \left(
2-\alpha \right) ,  \label{eq4}
\end{equation}
is universal, provided that
\begin{equation}
3\nu =2-\alpha ,  \label{eq5}
\end{equation}
holds. Moreover, additional universal relations include
\begin{equation}
\frac{A^{-}}{A^{+}}=R_{A},\ \frac{\xi _{x0}^{-}\xi _{y0}^{-}\xi _{z0}^{-}}{%
\xi _{x0}^{+}\xi _{y0}^{+}\xi _{z0}^{+}}=\frac{\xi _{x0}^{tr}\xi
_{y0}^{tr}\xi _{z0}^{tr}}{\xi _{x0}^{+}\xi _{y0}^{+}\xi
_{z0}^{+}}=R_{\xi }. \label{eq5b}
\end{equation}
Thus, the critical amplitudes are expected to differ from system
to system and to depend on the dopant concentration, the universal
combinations (\ref {eq2d}), (\ref{eq4}) and (\ref{eq5b}) should
hold for all cuprates and irrespective of the doping level, except
at the critical endpoints of the 3D-XY critical line.

A characteristic property of cuprate superconductors is their
anisotropy. \ In tetragonal systems, where $\xi _{a}=\xi _{b}=\xi
_{ab}$, it is defined as the ratio $\gamma =\xi _{ab}/\xi _{c}$,
of the correlation length parallel and perpendicular to the
ab-planes. Noting that according to Eqs.(\ref{eq2c}) and
(\ref{eq2g})
\begin{equation}
\frac{\Upsilon _{x}}{\Upsilon _{z}}=\frac{\xi _{z}^{tr}}{\xi
_{x}^{tr}}=\frac{\xi _{z}^{tr}\xi _{y}^{tr}}{\xi _{x}^{tr}\xi
_{y}^{tr}}=\frac{\lambda _{z}^{2}}{\lambda _{x}^{2}}=\left(
\frac{\xi _{x}^{-}}{\xi _{z}^{-}}\right) ^{2},  \label{eq6}
\end{equation}
holds, we obtain for $\gamma $ the relation
\begin{equation}
\gamma _{T_{c}}=\frac{\xi _{ab0}^{-}}{\xi _{c0}^{-}}=\frac{\lambda
_{c0}}{\lambda _{ab0}}.  \label{eq7}
\end{equation}
the universal relation (\ref{eq2d}) can then be rewritten in the
form

\begin{equation}
k_{B}T_{c}=\frac{\Phi _{0}^{2}}{16\pi ^{3}}\frac{\xi
_{ab,0}^{-}}{\lambda _{ab,0}^{2}\gamma _{T_{c}}}.  \label{eq8}
\end{equation}
Clearly, $T_{c}$, $\xi _{ab,0}^{-}$, $\lambda _{ab,0}$ and $\gamma
$ depend on the dopant concentration, but universality implies
that this relation applies at any finite temperature, irrespective
of the doping level. An other remarkable consequence follows from
the universal relation
\begin{equation}
k_{B}T_{c}^{3}=\left( \frac{\Phi _{0}^{2}}{16\pi ^{3}k_{B}}\right)
^{3}\frac{\left( R^{-}\right) ^{3}}{A^{-}\lambda _{x0}^{2}\lambda
_{y0}^{2}\lambda _{z0}^{2}}  \label{eq8a}
\end{equation}
which follows from Eqs.(\ref{eq2d}) and (\ref{eq4}). Indeed,
considering the effect of doping, substitution and pressure,
denoted by the variable y, we obtain
\begin{equation}
\frac{3}{T_{c}}\frac{dT_{c}}{dy}=-\frac{1}{A^{-}}\frac{dA^{-}}{dy}+\sum_{i}\lambda
_{i0}^{2}\frac{d\left( 1/\lambda _{i0}^{2}\right) }{dy}.
\label{eq8b}
\end{equation}
Thus, the effect of doping, substitution and pressure on
transition temperature, specific heat and penetration depths are
not independent, but related by this law.

In an applied magnetic field the singular part of the free energy
density adopts the scaling form\cite{book,tseuro}
\begin{equation}
f_{s}=-\frac{k_{B}TQ_{3}^{\pm }}{\xi _{x}^{\pm }\xi _{y}^{\pm }\xi
_{z}^{\pm }}G_{3}^{\pm }(\widetilde{z})\quad ,\quad G_{3}^{\pm
}(0)=1,  \label{eqc8}
\end{equation}
where
\begin{equation}
\widetilde{z}={{\frac{1}{\Phi _{0}}}\sqrt{H_{x}^{2}\xi _{y}^{2}\xi
_{z}^{2}+H_{y}^{2}\xi _{x}^{2}\xi _{z}^{2}+H_{z}^{2}\xi
_{x}^{2}\xi _{y}^{2}},}
\label{eqc9}
\end{equation}
and $G_{3}^{\pm }(\widetilde{z})$ is a universal scaling function
of its argument. Note that in the isotropic case, where $\xi =\xi
_{x}=\xi _{y}=\xi _{z}$, this scaling form is identical to that of
uniformly rotating $^{4}$He near the superfluid transition.
Magnetic field and rotation frequency are related by $H\rightarrow
m_{4}c\Omega /e$\cite{vinen}. In the presence of a magnetic field
and $T<T_{c}$ the correlations do no longer decay algebraically.
The Fourier transform of $S_{tr}\left( R\right) $ behaves for
small wavenumbers $q$ as
\begin{equation}
S_{tr}\left( q\right) \propto \frac{1}{q^{2}+1/\xi ^{2}}.
\label{eqd9}
\end{equation}
Supposing then that there is a phase transition in the (H,T)-plane
for $T<T_{c}$ the scaling function must have a singularity at some
value $\widetilde{z}_{c}$. Examples are the vortex melting and
vortex glass transition. Since the vortex melting transition is
first order, $\xi $ does not diverge but is bounded by
\begin{equation}
{{\frac{1}{\Phi _{0}}}\sqrt{H_{x}^{2}\xi _{y}^{2}\xi
_{z}^{2}+H_{y}^{2}\xi _{x}^{2}\xi _{z}^{2}+H_{z}^{2}\xi
_{x}^{2}\xi _{y}^{2}}}=\widetilde{z}_{m}. \label{eqe9}
\end{equation}
Invoking Eqs.(\ref{eq2}) we obtain for the first order transition
line with H $\Vert $c$\Vert z$ the expression,
\begin{equation}
H_{cm}=\frac{\Phi _{0}\widetilde{z}_{m}}{\xi _{a,0}\xi
_{b,0}}\left( \frac{T_{c}-T}{T_{c}}\right) ^{2\nu },\
\frac{T_{c}-T_{cm}}{T_{c}}=\left( \frac{\xi _{a,0}\xi
_{b,0}}{\widetilde{z}_{m}\Phi _{0}}\right) ^{1/2\nu }H_{c}^{1/2\nu
}\text{.}
\label{eqf9}
\end{equation}

In the interval $T_{cm}<T<T_{c}$ one expects a remnant of the zero
field specific heat singularity. Because the correlation length is
bounded, there is a magnetic field induced finite size effect. At
the melting transition the limiting length is $L^{2}=\left(
\widetilde{z}_{m}\Phi _{0}\right) /H_{c} $ (Eq.(\ref{eqe9})).
Close to $T_{c}$ on expects on dimensional grounds, $L^{2}\approx
\left( \Phi _{0}\right) /H_{c}$ to hold. Since the correlation
length cannot exceed $L$, the zero field singularity, i.e. in the
specific heat, is removed. As a remnant of this singularity, the
specific heat will also exhibit a maximum at $T_{p}$ which is
located below $T_{c}$ according to
\begin{equation}
\frac{T_{c}-T_{p}}{T_{c}}\approx \left( \frac{\xi _{a,0}\xi
_{b,0}}{\Phi _{0}}\right) ^{1/2\nu }H_{c}^{1/2\nu }.
\label{eqg9}
\end{equation}

\subsection{Evidence for finite temperature critical behavior}

\begin{figure}[tbp]
\centering
\includegraphics[totalheight=4cm]{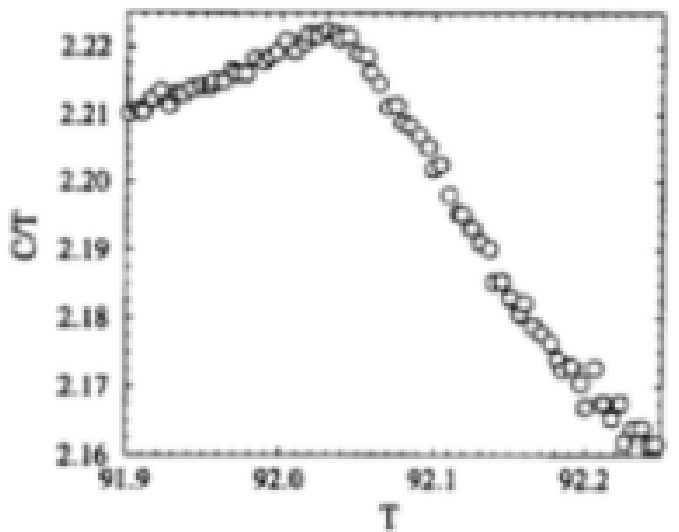}
\caption{Specific heat coefficient C/T $\left( mJ/\left(
gK^{2}\right) \right) $ versus T $\left( K\right) $ of
$YBa_{2}Cu_{3}O_{7-\protect\delta }$ ( sample YBCO3
\protect\cite{chara}). The two arrows mark $T_{c}\approx 92.12 K$
and $T_{P}\approx 91.98 K$ , respectively.} \label{fig8}
\end{figure}

Provided that the thermal critical behavior is
fluctuation-dominated (i.e. non-mean-field) and the fluctuations
of the vector potential can be neglected, cuprate superconductors
fall into the 3D-XY universality class. We have seen that the
universality class to which a given system belongs is not only
characterized by its critical exponents but also by various
critical-point amplitude combinations. The implications include:
(i) The universal relations hold irrespective of the dopant
concentration and material; (ii) Given the nonuniversal critical
amplitudes of the correlation lengths, $\xi _{i,0}^{\pm }$, and
the universal scaling function $G_{3}^{\pm }(\widetilde{z})$,
universal properties can be derived from the singular part of the
free energy close to the zero field transition. These properties
include the specific heat, magnetic torque, diamagnetic
susceptibility, melting line, etc. Although there is mounting
evidence for 3D-XY-universality in cuprates\cite
{hoferdis,book,ariosa,tskel,hubbard,kamal,pasler,tseuro,houston},
it should be kept in mind, that evidence for power laws and
scaling should properly consist of experimental data that covers
several decades of the parameters. In practice, there are
inhomogeneities and cuprates are homogeneous over a finite length
$L$ only. In this case, the actual correlation length $\xi
(t)\propto |t|^{-\nu }$ cannot grow beyond $L$ as $t\rightarrow
0$, and the transition appears rounded. Due to this finite size
effect, the specific heat peak occurs at a temperature $T_{P}$
shifted from the homogeneous system by an amount $L^{-1/\nu }$,
and the magnitude of the peak located at temperature $T_{P}$
scales as $L^{\alpha /\nu }$. To quantify this point we show in
Fig.\ref{fig8} the measured specific heat coefficient of
YBa$_{2}$Cu$_{3}$O$_{7-\delta }$\cite{chara}. The rounding and the
shape of the specific heat coefficient clearly exhibits the
characteristic behavior of a system in confined dimensions, i.e.
rod or cube shaped inhomogeneities\cite{schultka}.

\begin{figure}[tbp]
\centering
\includegraphics[totalheight=5cm]{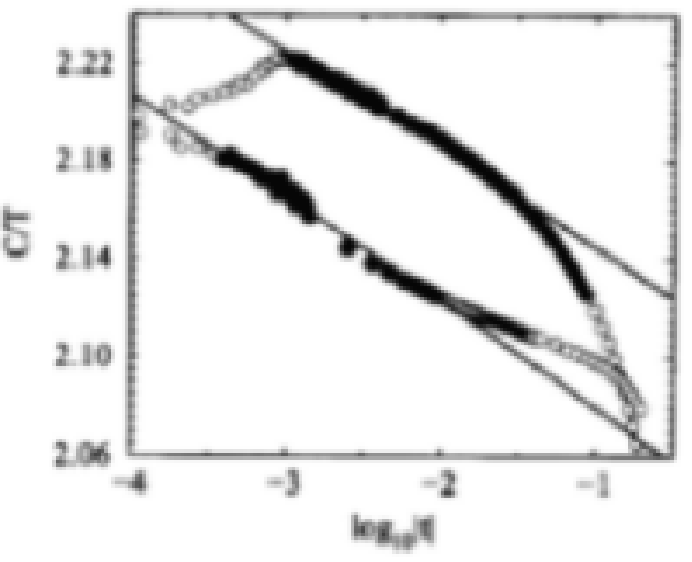}
\caption{Specific heat coefficient C/T $\left( mJ/\left(
gK^{2}\right) \right) $ versus log$_{10}\left| t\right| $ for
$YBa_{2}Cu_{3}O_{7-\protect\delta }$ ( sample YBCO3
\protect\cite{chara}) for $T_{c}=92.12 K$.} \label{fig9}
\end{figure}
A finite-size scaling analysis\cite{book} reveals inhomogeneities
with a characteristic length scale ranging from $300$ to $400$ A,
in the YBa$_{2}$Cu$_{3}$O$_{7-\delta }$ samples YBCO3, UBC2 and
UBC \cite{chara}. Note that recent measurements by a variety of
techniques suggest that superconductivity is not homogeneous in
cuprates\cite{bozin,lang,hunt}.For this reason, deviations from
$3D$-$XY$ critical behavior around $T_{p}$ do not signal the
failure of $3D$-$XY$ universality, as previously
claimed\cite{chara}, but reflect a finite-size effect at work.
Indeed from Fig.\ref{fig9} it is seen that the finite-size effect
makes it impossible to enter the asymptotic critical regime. To
set the scale we note that in the $\lambda $-transition of
$^{4}$He the critical properties can be probed down to $\left|
t\right| =10^{-9}$\cite{sinsgaas,mehta}. In Fig.\ref {fig8} we
marked the intermediate regime where consistency with the
3D-XY-critical behavior, ${C}/{T}=\widetilde{A}^{\pm }10^{-\alpha
\log _{10}|t|}+\widetilde{B}^{\pm }$ for $\alpha =-0.013$ and
$\widetilde{A}^{+}/\widetilde{A}^{-}=1.07$, can be observed in
terms of full circles. The upper branch corresponds to $T<T_{c}$ \
and the lower one to $T>T_{c}$. The open circles closer to $T_{c}$
correspond to the finite-size affected region, while further away
the temperature dependence of the background, usually attributed
to phonons, becomes significant. Hence, due to the finite size
effect and the temperature dependence of the background the
intermediate regime is bounded by the temperature region where the
data depicted in Fig. \ref{fig9} fall nearly on straight lines. In
this context it should be kept in mind that the effect of disorder
and inhomogeneities is quite different. Since the critical
exponent $\alpha $ of the specific heat is negative at the 3D-XY
transition, Harris criterion implies that disorder is irrelevant
so that the critical behavior will be that of the pure system
\cite{ma}. To provide quantitative evidence for 3D-XY universality
in this regime, we invoke the universal relations (\ref{eq2d}) and
(\ref{eq4}) to calculate $T_{c}$ from \ the critical amplitudes of
specific heat and penetration depth. Using $A^{+}=8.4\ 10^{20}$
$cm^{3}$, derived from the data shown in Fig.\ref{fig9} for sample
YBCO3 with\ $T_{c}=92.12 K$, $\lambda _{a,0}=1153A,\ \lambda
_{b,0}=968A$ and $\ \lambda _{c,0}=8705A$ \ , derived from
magnetic torque measurements on a sample with $T_{c}=91.7 K$ \cite
{tseuro}, together with the universal numbers $A^{+}/A^{-}=1.07$
and $R^{-}\approx 0.59$\cite{book}, we obtain $T_{c}=88.2 K$ .
Hence, the universal 3D-XY-relations (\ref{eq2d}) and (\ref{eq4})
are remarkably well satisfied.

\begin{figure}[tbp]
\centering
\includegraphics[totalheight=5cm]{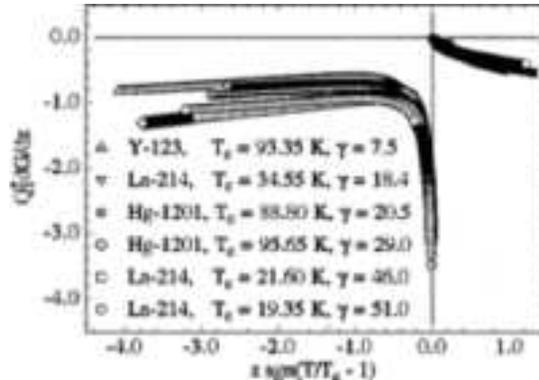}
\caption{Scaling function $dG_{3}^{\pm
}(\widetilde{z})/d\widetilde{z}$
derived from \ the angular dependence of the magnetic torque for YBa$_{2}$Cu$%
_{3}$O$_{6.93}$, La$_{1.854}$Sr$_{0.146}$CuO$_{4}$,
HgBa$_{2}$CuO$_{4.108}$,
HgBa$_{2}$CuO$_{4.096}$, La$_{1.914}$Sr$_{0.086}$CuO$_{4}$ and La$_{1.920}$Sr%
$_{0.080}$CuO$_{4}$. Taken from \protect\cite{hoferdis}.}
\label{fig10}
\end{figure}

Another difficulty in observing 3D-XY critical properties stems
from the fact that most cuprates are highly anisotropic. A
convenient measure of the anisotropy is $\gamma _{T_{c}}$
(Eq.(\ref{eq7})), which depends on the dopant concentration (see
Figs.\ref{fig1} and \ref{fig2}). Although the strength of thermal
fluctuations grows with increasing $\gamma $, they become
essential 2D slightly away from $T_{c}$. Accordingly, the 3D-XY -
critical regime shrinks and the corrections to scaling become
significant. To document this point we depicted in Fig.\ref{fig10}
estimates for the derivative of the universal scaling function
$G_{3}^{\pm }(\widetilde{z})$ derived from magnetic torque
measurements\cite{hoferdis}. Even though the qualitative behavior
is the same for all samples, the deviations increase with rising
$\gamma _{T_{c}}$. This systematics cannot be attributed to the
experimental uncertainties of about 40\%. It is more likely that
it reflects the reduction of the temperature regime where
3D-fluctuation dominates so that corrections to scaling become
important\cite{book}. In this view it is
clear that due its moderate anisotropy\cite{book,tseuro}, optimally doped YBa%
$_{2}$Cu$_{3}$O$_{7-\delta }$ is particularly suited to observe
and check the consistency with 3D-XY critical behavior. In
contrast to this, in highly anisotropic cuprates like
Bi$_{2}$Sr$_{2}$CaCu$_{2}$O$_{8-\delta }$, it will be difficult to
enter the regime where 3D fluctuations dominate \cite {junodbi}.
Nevertheless, since the critical behavior of the charged fixed
point is the only alternative left, it becomes clear that even the
intermediate critical behavior of highly anisotropic cuprates like Bi$_{2}$Sr%
$_{2}$CaCu$_{2}$O$_{8-\delta }$ falls into the 3D-XY universality
class. fluctuations grows with increasing $\gamma $, they become
essential 2D slightly away from $T_{c}$. Accordingly, the 3D-XY -
critical regime shrinks and the corrections to scaling become
significant. To document this point we depicted in Fig.\ref{fig10}
estimates for the derivative of the universal scaling function
$G_{3}^{\pm }(\widetilde{z})$ derived from magnetic torque
measurements\cite{hoferdis}. Even though the qualitative behavior
is the same for all samples, the deviations increase with rising
$\gamma _{T_{c}}$. This systematics cannot be attributed to the
experimental uncertainties of about 40\%. It is more likely that
it reflects the reduction of the temperature regime where
3D-fluctuation dominates so that corrections to scaling become
important\cite{book}. In this view it is
clear that due its moderate anisotropy\cite{book,tseuro}, optimally doped YBa%
$_{2}$Cu$_{3}$O$_{7-\delta }$ is particularly suited to observe
and check the consistency with 3D-XY critical behavior. In
contrast to this, in highly anisotropic cuprates like
Bi$_{2}$Sr$_{2}$CaCu$_{2}$O$_{8-\delta }$, it will be difficult to
enter the regime where 3D fluctuations dominate \cite {junodbi}.
Nevertheless, since the critical behavior of the charged fixed
point is the only alternative left, it becomes clear that even the
intermediate critical behavior of highly anisotropic cuprates like Bi$_{2}$Sr%
$_{2}$CaCu$_{2}$O$_{8-\delta }$ falls into the 3D-XY universality
class.

\begin{figure}[tbp]
\centering
\includegraphics[totalheight=5cm]{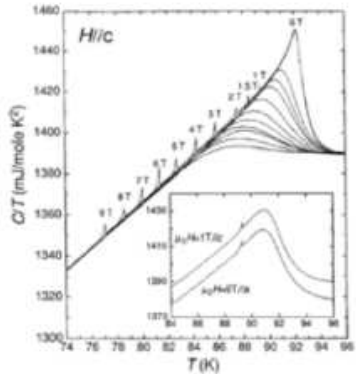}
\caption{Temperature dependence of the specific heat coefficient
for various applied magnetic fields $\left( H\Vert c\right) $ of
an untwinned YBa$_{2}$Cu$_{3}$O$_{7-\protect\delta }$ single
crystal. The numbers on the top of the peak like features denote
the applied field strength. The inset shows representative data
for $H=1T$ $\left( H\Vert c\right) $ and $H=8T$ $\left( H\Vert
(a,b)\right) $, data shifted vertically by 10 mJ/molK$^{2}$. Taken
from \protect\cite{schillingm}.} \label{fig11}
\end{figure}

The melting transition of the vortex lattice was discovered in 1993 in Bi$%
_{2.15}$Sr$_{1.85}$CaCu$_{2}$O$_{8-\delta }$ using the $\mu $SR
technique\cite{lee}. An anomaly attributed to this transition was
observed in specific heat measurements of
YBa$_{2}$Cu$_{3}$O$_{7-\delta }$\cite{schillingm}. In
Fig.\ref{fig11} we depicted the temperature dependence of the
specific heat coefficient of an untwinned
YBa$_{2}$Cu$_{3}$O$_{7-\delta }$ single crystal for various
applied fields H$\Vert $c. Of particular interest in this context
is the small anomaly below the main peak, marked by the strength
of the applied field. It is attributed to the vortex melting
transition. Evidence for the first order nature of the transition
stems from magnetization measurements, revealing a jump at
$H_{m}$\cite{liang}, which signals the singularity in the scaling
function $G_{3}^{\pm }(\widetilde{z})$ at
$\widetilde{z}=\widetilde{z}_{m}$ (Eq.(\ref {eqc8})).

\begin{figure}[tbp]
\centering
\includegraphics[totalheight=5cm]{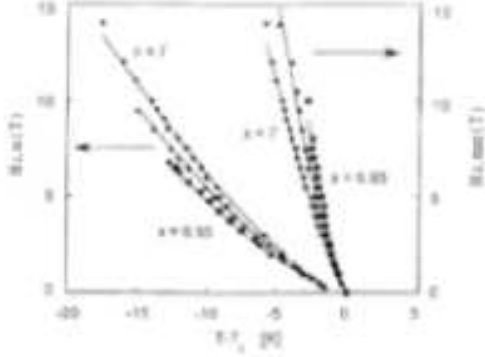}
\caption{Melting line $H_{c,m}$ and $H_{c,\max }$ versus $T-T_{c}$
for a $YBa_{2}CuO_{7-\protect\delta }$ single crystal with
$\protect\delta =0$ ($\blacklozenge $), $\protect\delta =0.03$
($\blacktriangle $) and $\protect\delta =0.053$ ($\bullet $) for
H$\Vert $c. Taken from \protect\cite{roulinm}. The curves
correspond to $H_{cm}=A_{cm}\left( \frac{T_{c}-T}{T_{c}}\right)
^{2\protect\nu }$ and $H_{cp}=A_{cp}\left(
\frac{T_{c}-T}{T_{c}}\right) ^{2\protect\nu }$ with$\ \protect\nu
=2/3$.}
\label{fig12}
\end{figure}

Due to the first order nature of the transition, the correlation
lengths remain bounded, so that the melting line $H_{cm}$ and the
temperature $T_{P}$, where the broad peak in the specific heat
coefficient adopts its maximum value, should scale according to
Eqs.(\ref{eqf9}) and (\ref{eqg9}). A glance to Fig.\ref{fig12}
shows that this behavior is experimentally well confirmed. Note
that the variation of the amplitudes $A_{cm}$ and $A_{cp}$ is due
to the doping dependence of $\xi _{a,0}\xi _{b,0} $ ( see
Eqs.(\ref{eqf9}) and (\ref{eqg9})). From Eq.(\ref{eqe9}) it is
seen that the melting line also yields useful information on the
anisotropy. As an example we consider the angular dependence in
the (c,a) plane. Eq.(\ref{eqe9}) yields
\begin{equation}
H_{m}\left( \delta \right) =H_{m}\left( \delta =0\right) \gamma
_{ca}\left( \sin ^{2}\left( \delta \right) +\gamma _{ca}^{2}\cos
^{2}\left( \delta \right) \right) ^{-1/2},\text{ }\gamma
_{ca}=\frac{\xi _{a}}{\xi _{c}} ,H_{m}\left( \delta =0\right)
\propto \left( 1-T/T_{c}\right) ^{2\nu }.
\label{eqh9}
\end{equation}

\begin{figure}[tbp]
\centering
\includegraphics[totalheight=5cm]{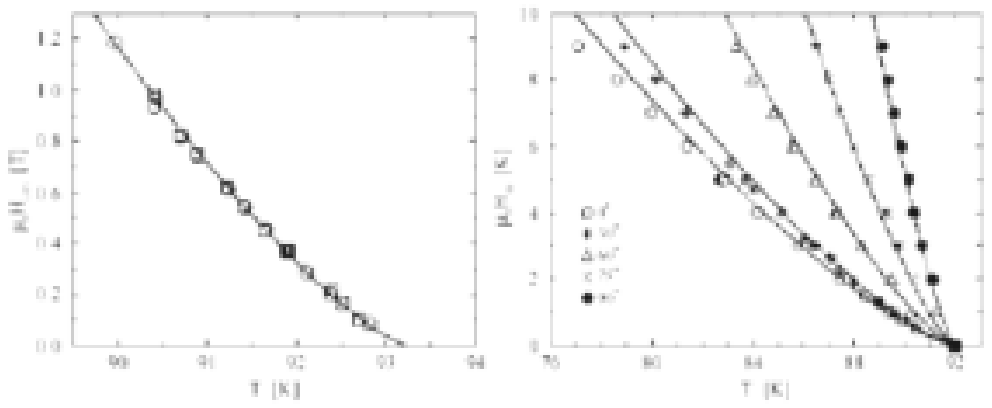}
\caption{Left panel: Melting line for\
YBa$_{2}$Cu$_{3}$O$_{7-\protect\delta }$ with $T_{c}=93.25 K$
derived from field $\left( \bigcirc \right) $- and
angular-dependent $\left( \square \right) $ torque curves. The
solid line corresponds to $H_{m}\propto \left( 1-T/T_{c}\right)
^{2\protect\nu }$ with $\protect\nu =2/3$. Data taken from
\protect\cite{willeminm}. Right panel: Melting lines $H_{m}\left(
\protect\delta ,T\right) $ for\
YBa$_{2}$Cu$_{3}$O$_{7-\protect\delta }$ with $T_{c}=92 K$,
detected by a direct measurement of the entropy change. The
symbols denote experimental data taken at different magnetic field
orientations in the (c,a) plane, measured by the angle
$\protect\delta $\protect\cite{schillingmang}. The solid lines are
Eq.(\ref {eqh9}) with $\protect\gamma _{ca}=7.6$ and $\protect\nu
=2/3$.} \label{fig13}
\end{figure}

From Fig.\ref{fig13} it is seen that this behavior is well
confirmed. In this context the question arises, whether or not
phase coherence and with that superfluidity persists above the
melting line. Recently, numerical simulations revealed that the
vortex liquid is incoherent, i.e. phase coherence is destroyed in
all directions, including the direction of the applied magnetic
field, as soon as the vortex lattice melts\cite{hu,sudbo}.

To summarize, there is considerable evidence that the intermediate
finite temperature critical behavior of cuprate superconductors,
when attained, is equivalent to that of superfluid helium.
Moreover, we saw that finite size scaling is a powerful tool to
identify and characterize inhomogeneities.

\section{Quantum critical behavior and crossover phenomena}

\subsection{Sketch of the scaling predictions}

Given the empirical phase transition line $T_{c}\left( x\right) $
or surface $T_{c}\left( x,y\right) $ with critical endpoints or
lines (see Figs.\ref {fig1} and \ref{fig3}) doping and
substitution tuned quantum phase transitions can be expected. To
invoke and sketch the scaling theory of quantum critical phenomena
we define $\delta $, measuring the relative distance from quantum
critical points, in terms of
\begin{equation}
\delta =\left\{
\begin{array}{c}
\ \ \ \ \ y=0\ \ \ \ \ \ \ \ \ \ \ \ \ \ \ \ \ \ \ \ \ :\delta
=\left(
x-x_{u}\right) /x_{u} \\
\ \ \ \ \ y=0\ \ \ \ \ \ \ \ \ \ \ \ \ \ \ \ \ \ \ \ \ \ :\delta
=\left(
x_{o}-x\right) /x_{o} \\
y\neq 0,\ \ x_{u}\leq x\leq x_{o}\ \ \ \ \ \ \ \ \ \ \ \ \ :\delta
=\left( y_{c}\left( x\right) -y\right) /y_{c}\left( x\right)
\end{array}
\right. .  \label{eq9}
\end{equation}
At $T=0$ and close to quantum criticality one has two kind of
correlation lengths\cite{book,kim}. The usual spatial correlation
length in direction $i$
\begin{equation}
\overline{\xi }_{i}^{-}=\overline{\xi }_{i,0}^{-}\delta
^{-\overline{\nu }} \label{eq10}
\end{equation}
and the temporal one
\[
\overline{\xi }_{\tau }^{-}=\overline{\xi }_{\tau ,0}^{-}\delta
^{-\overline{ \nu }_{\tau }},
\]
where the dynamic critical exponent is defined as the ratio
\begin{equation}
z=\frac{\overline{\nu }_{\tau }}{\overline{\nu }}.  \label{eq11}
\end{equation}
The singular part of the free energy density adopts then the
scaling form \cite{book,kim,klosters}

\begin{equation}
f_{s}\left( \delta ,T\right) =\overline{Q}_{D}\left( \overline{\xi
}_{\tau ,0}^{-}\prod\limits_{i=1}^{D}\overline{\xi
}_{i,0}^{-}\right) ^{-1}\delta ^{ \overline{\nu }\left( D+z\right)
}F_{D}\left( \overline{y}\right) ,\ \ \
\overline{y}=k_{B}T\overline{\xi }_{\tau }=k_{B}T\overline{\xi
}_{\tau ,0}\delta ^{-z\overline{\nu }}, \label{eq12}
\end{equation}
where $F_{D}\left( \overline{y}\right) $ with $F_{D}\left(
\overline{y} =0\right) =1$ is a universal scaling function and
$Q_{D}$ a universal constant. Another quantity of interest is the
helicity modulus which adopts the scaling form
\cite{book,klosters}
\begin{equation}
\Upsilon _{i}^{D}\left( \delta ,T\right) =\overline{Q}_{D}\left(
\overline{\xi }_{i,0}^{-}\right) ^{2}\left( \overline{\xi }_{\tau
,0}^{-}\prod\limits_{i=1}^{D}\overline{\xi }_{i,0}^{-}\right)
^{-1}\delta ^{\overline{\nu }\left( D-2+z\right) }Y_{D}\left(
\overline{y}\right) , \label{eq13}
\end{equation}
where $Y_{D}\left( \overline{y}\right) $ with $Y_{D}\left(
y=0\right) =1$ is a universal scaling function of its argument. As
a first application we consider a line of finite temperature
transitions $T_{c}\left( \delta \right) $ ending at a quantum
critical point at $T=0$ and $\delta =0$. The scaling forms then
require that
\begin{equation}
k_{B}T_{c}=\frac{\overline{y}_{c}}{\overline{\xi }_{\tau
,0}^{-}}\delta ^{z \overline{\nu }},  \label{eq14}
\end{equation}
where $y_{c}$ is the universal value of the scaling function
argument at which the scaling functions exhibit a singularity at
finite temperature. Combining Eqs.(\ref{eq13}) and (\ref{eq14}) we
obtain in $D=2$
\begin{equation}
\frac{k_{B}T_{c}}{\Upsilon _{ab}^{D=2}\left( \delta ,0\right)
}=\overline{R} _{2},\ \ \
\overline{R}_{2}=\frac{\overline{y}_{c}}{Y_{2}\left( \overline{y}
_{c}\right) \overline{Q}_{2}},  \label{eq15}
\end{equation}
yielding the universal relation
\begin{equation}
T_{c}\lambda _{ab}^{2}\left( 0\right) =\frac{\Phi
_{0}^{2}\overline{R}_{2}}{ 16\pi ^{3}k_{B}}d_{s},  \label{eq16}
\end{equation}
between transition temperature and zero temperature in-plane
penetration depth. $d_{s}$ denotes the thickness of the
superconducting slab and $ \overline{R}_{2}$ is a universal
dimensionless constant. Analogously, in $ D=3 $ Eqs.(\ref{eq13})
and (\ref{eq14}) yield
\begin{equation}
\frac{k_{B}T_{c}}{\Upsilon _{ab}^{D=3}\left( \delta ,0\right)
}=\overline{R}_{3}\overline{\xi }_{c}^{-},\ \ \
\overline{R}_{3}=\frac{\overline{y}_{c}}{Y_{3}\left(
\overline{y}_{c}\right) \overline{Q}_{3}}, \label{eq17}
\end{equation}
so that $T_{c}$, $\lambda _{ab}^{2}\left( 0\right) $,
$\overline{\xi }_{c}^{-}$ and $\overline{\xi }_{ab}^{-}$ are
related by
\begin{equation}
T_{c}\lambda _{ab}^{2}\left( 0\right) =\frac{\Phi _{0}^{2}\ \overline{R}_{3}%
}{16\pi ^{3}k_{B}}\overline{\xi }_{c}^{-}=\frac{\Phi _{0}^{2}\
\overline{R}_{3}}{16\pi ^{3}k_{B}}\frac{\overline{\xi
}_{ab}^{-}}{\gamma _{T=0}}. \label{eq18}
\end{equation}
This is just the quantum analogy of the universal finite
temperature relation (\ref{eq8}). When there is an anisotropy
tuned 3D-2D crossover, where $\gamma _{T=0}\rightarrow \infty $,
matching of Eqs.(\ref{eq16}) and (\ref{eq18}) requires that
\begin{equation}
\overline{R}_{3}\overline{\xi
}_{c}^{-}=\overline{R}_{3}\frac{\overline{\xi }_{ab}^{-}}{\gamma
_{T=0}}=\overline{R}_{2}d_{s}.  \label{eq19}
\end{equation}
Noting then that the universal relation (\ref{eq18}) applies close
to both, the 2D-QSI and 3D-QSN transition it is expected to
provide useful information on the correlation lengths (
$\overline{\xi }_{c}^{-}$, $ \overline{\xi }_{ab}^{-}$), given
experimental data for $T_{c}\lambda _{ab}^{2}\left( 0\right) $ and
$\gamma _{T=0}$.

Moreover, useful scaling relations for the specific heat
coefficient $\gamma _{c}$ are readily derived from the singular
part of the free energy density (\ref{eq12}), namely:
\begin{equation}
\left. \gamma _{c}\right| _{T=0}=\left. \frac{c}{T}\right|
_{T=0}=\left. \frac{\partial ^{2}f_{s}}{\partial T^{2}}\right|
_{T=0}\propto \delta ^{\overline{\nu }\left( D-z\right) }\propto
T_{c}^{\left( D-z\right) /z}\propto H_{c}^{\left( D-z\right) /2},
\label{eq20}
\end{equation}
since $H_{c}$ scales as $H_{c}\propto \left( \overline{\xi
}_{ab}^{-}\right) ^{-2}$\cite{book}, when applied parallel to the
c-axis, and
\begin{equation}
\left. \frac{d\gamma _{c}}{dT}\right| _{T=0}=\left.
\frac{d}{dT}\left( \frac{c}{T}\right) \right| _{T=0}=\left.
\frac{\partial ^{3}f_{s}}{\partial T^{3}}\right| _{T=0}\propto
\delta ^{\overline{\nu }\left( D-2z\right) }\propto T_{c}^{\left(
D-2z\right) /z}.  \label{eq21}
\end{equation}
Finally, considering a 2D quantum critical points resulting from a
3D-2D crossover in the ground state, Eq.(\ref{eq19}) implies that
close to 2D quantum criticality the anisotropy diverges as
\begin{equation}
\gamma _{T=0}=\frac{\overline{\xi }_{ab}^{-}}{\overline{\xi
}_{c}^{-}}=\frac{\overline{R}_{3}\overline{\xi
}_{ab,0}^{-}}{\overline{R}_{2}d_{s}}\delta ^{-\overline{\nu
}}\propto T_{c}^{-1/z}.  \label{eq22}
\end{equation}
Combining the scaling relations (\ref{eq14}), (\ref{eq16}),
(\ref{eq19}) and (\ref{eq22}) a 2D quantum critical point
resulting from a 3D-2D crossover is then characterized by
\begin{equation}
T_{c}\propto {\lambda _{ab}^{-2}}\left( {0}\right) \propto
n_{s}^{\square }\left( 0\right) \propto \gamma _{T=0}^{-z}\propto
\delta ^{z\overline{\nu }},\ \ \ \left. \gamma _{c}\right|
_{T=0}\propto T_{c}^{\left( 2-z\right) /z}\propto H_{c}^{\left(
2-z\right) /2}.  \label{eq23}
\end{equation}
$n_{s}^{\square }\left( 0\right) =\left( d_{s}/{\lambda
_{ab}^{2}(0)}\right) $ denotes aerial superfluid density. In
particular it relates the superconducting properties to the
anisotropy parameter, fixing the dimensionality of the system. It
reveals that an anisotropy driven 3D-2D crossover destroys
superconductivity even in the ground state.

From Eq.(\ref{eq18}), rewritten in the form
\begin{equation}
\overline{\xi }_{c}^{-}=\frac{16\pi ^{3}k_{B}}{\Phi _{0}^{2}\
\overline{R}_{3}}T_{c}\lambda _{ab}^{2}\left( 0\right) ,\ \
\overline{\xi }_{ab}^{-}=\frac{16\pi ^{3}k_{B}}{\Phi _{0}^{2}\
\overline{R}_{3}}T_{c}\lambda _{ab}^{2}\left( 0\right) \gamma
_{T=0},  \label{eq24}
\end{equation}
it is seen that $T_{c}\lambda _{ab}^{2}\left( 0\right) $ and
$T_{c}\lambda _{ab}^{2}\left( 0\right) \gamma _{T=0}$ are
appropriate indicators for the
occurrence of quantum phase transitions. Close to 3D-QSN criticality $%
\overline{\xi }_{c}^{-}$ and $\overline{\xi }_{ab}^{-}$ diverge
according \ Eq.(\ref{eq10}) as
\begin{equation}
\text{3D-QSN}:\overline{\xi }_{c}^{-}\propto \ \overline{\xi }%
_{ab}^{-}\propto \delta ^{-\overline{\nu }}\propto T_{c}^{-1/z},
\label{eq25}
\end{equation}
because $\gamma _{T=0}$ remains finite. This differs from the
2D-QSI transition, where according to relation (\ref{eq22})
\begin{equation}
\text{2D-QSI}:\overline{\xi }_{c}^{-}=\frac{\overline{R}_{2}}{\overline{R}%
_{3}}d_{s},\ \ \overline{\xi }_{ab}^{-}=\frac{\overline{R}_{2}}{\overline{R}%
_{3}}d_{s}\gamma _{T=0}\propto \delta ^{-\overline{\nu }}\propto
T_{c}^{-1/z},  \label{eq26}
\end{equation}
applies. Here $\overline{\xi }_{c}^{-}$ tends to a finite value,
proportional to the thickness $d_{s}$ of the sheets. valid for a
2D-XY - QSI transition. $\rho _{0}$ and $\rho _{0c}$ denote the
residual and critical residual sheet resistivity.

Next we turn to the critical behavior of the conductivity. In the
normal
state and close to a 3D-XY critical point, the DC conductivities, parallel ($%
\sigma _{ab}^{DC}$) and perpendicular ($\sigma _{c}^{DC}$) to the
layers, scale as\cite{book}
\begin{equation}
\sigma _{ab}^{DC}\propto \frac{\xi _{\tau }}{\xi _{c}},\ \sigma
_{c}^{DC}\propto \frac{\xi _{c}\xi _{\tau }}{\xi _{ab}^{2}},
\label{eq27}
\end{equation}
where $\xi _{\tau }$ is the correlation length associated with the
finite temperature critical dynamics. At $T_{c}$ the ratio is then
simply given by the anisotropy
\begin{equation}
\frac{\sigma _{c}^{DC}}{\sigma _{ab}^{DC}}=\left( \frac{\xi _{c}}{\xi _{ab}}%
\right) ^{2}=\left( \frac{1}{\gamma _{T_{c}}}\right) ^{2}.
\label{eq28}
\end{equation}
Approaching 2D-QSI criticality, the scaling relation (\ref{eq23})
implies
\begin{equation}
\gamma _{T_{c}}=\gamma _{T_{c},0}\delta ^{-\overline{\nu }}
\label{eq29}
\end{equation}
Another essential property of this critical point is that for any finite $%
T_{c}$ the in-plane areal conductivity is always larger
than\cite{book,kim}
\begin{equation}
\sigma _{ab}^{DC}d_{s}=\sigma _{0}\frac{4e^{2}}{h}.  \label{eq30}
\end{equation}
This follows from the fact that close to 2D-QSI the in-plane
resistivity adopts the scaling form\cite{book,kim}
\begin{equation}
\rho _{ab}=\frac{h}{4e^{2}\sigma _{0}}F\left( \overline{y}\right)
,\ \ \overline{y}=k_{B}T\overline{\xi }_{\tau }.  \label{eqb30}
\end{equation}
Thus, close to a 2D-QSI transition the normal state resistivity
$\rho
_{c}^{DC}\left( T_{c}^{+}\right) =1/\sigma _{c}^{DC}\left( T_{c}^{+}\right) $%
, evaluated close but slightly above $T_{c}$, is predicted to
diverge as
\begin{equation}
\rho _{c}^{DC}\left( T_{c}^{+}\right) =\gamma _{T_{c}}^{2}\frac{h}{%
4e^{2}\sigma _{0}}=\gamma _{T_{c},0}^{2}\delta ^{-2\overline{\nu }}\frac{h}{%
4e^{2}\sigma _{0}}.  \label{eq31}
\end{equation}
$F\left( y\right) $ is a scaling function of its argument and
$F\left( 0\right) =1$. This behavior uncovers the 3D-2D crossover
associated with the flow to 2D-QSI criticality in the normal
state. To establish a relation between normal and superconducting
properties, we express $\delta $ in terms of $\lambda _{c}\left(
0\right) $. Using Eq.(\ref{eq23}) we obtain
\begin{equation}
\lambda _{c}\left( 0\right) =\Omega _{s}\left( \sigma
_{c}^{DC}\left( T_{c}^{+}\right) \right) ^{-\left( 2+z\right)
/4},\ \ \ \Omega _{s}=\gamma _{0,0}\lambda _{ab,0}\left( 0\right)
\left( \frac{4e^{2}\sigma _{0}}{h\gamma _{T_{c},0}^{2}}\right)
^{\left( 2+z\right) /4},  \label{eq32}
\end{equation}
where
\begin{equation}
\gamma _{T}=\gamma _{T,0}\delta ^{-\overline{\nu }},\ \ \lambda
_{ab,0}\left( 0\right) =\lambda _{ab,0}\left( 0\right) \delta ^{-z\overline{%
\nu }/2},  \label{eq33}
\end{equation}
are the zero temperature critical amplitudes. The scaling relation
(\ref {eq32}) differs from the mean-field prediction for bulk
superconductors in the dirty limit \cite{tinkham} and layered BCS
superconductors, treated as weakly coupled Josephson
junction\cite{bulaevski,ambegaokar,deutscher}:
\begin{equation}
\lambda _{c}\left( 0\right) =\Omega _{s}\left( \sigma
_{c}^{DC}\left( T_{c}^{+}\right) \right) ^{-1/2},  \label{eq34}
\end{equation}
where
\begin{equation}
\Omega _{s}=\left( \frac{\hbar c^{2}}{4\pi ^{2}\Delta \left( 0\right) }%
\right) ^{1/2},\text{ \ }\Delta \left( 0\right) =1.76k_{B}T_{c}
\label{eq35}
\end{equation}
and $\Delta \left( 0\right) $ denotes the zero temperature energy
gap.

Approaching 3D-QSN criticality, the finite temperature relations (\ref{eq27}%
) and (\ref{eq28}) still apply, but both, $\rho _{c}^{DC}$ and
$\rho _{ab}^{DC}$ diverge at $T_{c}$, while the anisotropy $\gamma
_{T_{c}}$ remains finite. For this reason $\xi _{\tau }$, the
correlation length associated with the finite temperature critical
dynamics, cannot be eliminated. Nevertheless, since $\xi _{\tau }$
scales as, $\xi _{\tau }\propto \xi _{ab}^{z_{cl}}$, where
$z_{cl}$ is the critical exponent of the finite temperature
dynamics, we obtain from Eq.(\ref{eq27}) the relation
\begin{equation}
\sigma _{c}^{DC}\propto \frac{\xi _{c}\xi _{\tau }}{\xi
_{ab}^{2}}\propto \frac{\xi _{\tau }}{\gamma _{T_{c}}\xi
_{ab}}\propto \xi _{ab}^{z_{cl\ }-1}\propto \xi _{ab,0}^{z_{cl\
}-1}\left| t\right| ^{-\nu \left( z_{cl\ }-1\right) },
\label{eq36}
\end{equation}
valid close to 3D-XY critical points. Since close to a 3D-QSN
criticality, the doping dependence of the finite temperature
critical amplitude $\xi _{ab,0}$ is given by $\overline{\xi
}_{ab}$ ( see Eqs.(\ref{eq8}) and (\ref {eq18})), we finally
obtain with $\ \lambda _{c}\left( 0\right) \propto \delta
^{-\frac{\overline{\nu }}{2}\left( 1+z\right) }$ (Eq.(\ref{eq13}))
the relationship between $\lambda _{c}\left( 0\right) $ and
$\sigma _{c}\left( T_{c}^{+}\right) $ the relationship
\begin{equation}
\lambda _{c}\left( 0\right) \propto \left( \sigma _{c}\left(
T_{c}^{+}\right) \right) ^{\frac{1+z}{2\left( z_{cl}-1\right) }},
\label{eq37}
\end{equation}
characterizing the flow to a 3D-QSN critical point in the $\left(
\lambda _{c}\left( 0\right) ,\sigma _{c}\left( T_{c}^{+}\right)
\right) $ plane.

Noting that the in-plane resistivity tends close to 2D-QSI
criticality to a fixed value (Eq.(\ref{eqb30})), $\rho _{ab}$ can
also be used as a parameter measuring the distance from the
critical point. This is achieved by setting $y=k_{B}T\overline{\xi
}_{\tau }\propto T\delta ^{-z\overline{\nu }}\propto T\left(
\left( \rho _{ab}^{0}-\rho _{ab}\right) /\rho _{ab}^{0}\right)
^{-z\overline{\nu }}$. Given then a transition line $T_{c}\left(
\delta \right) $ ending at the 2D-QSN critical point the scaling
form (\ref{eqb30}) requires that
\begin{equation}
T_{c}\propto \left( \frac{\rho _{ab}^{0}-\rho _{ab}}{\rho
_{ab}^{0}}\right) ^{z\ \overline{\nu }},\ \rho
_{ab}^{0}=\frac{h}{4e^{2}\sigma _{0}} \label{eq38}
\end{equation}
because the scaling function $F\left( y\right) $ exhibits a
singularity at $y_{c}$, signaling the finite temperature
transition line.

\subsection{Evidence for chemically doping tuned quantum phase transitions}

The empirical correlations between $T_{c}$, dopant concentration
$x$ and anisotropy $\gamma _{T}$ (Eqs.(\ref{eq1a})-(\ref{eq1c}))
clearly point to the existence of quantum critical endpoints. A
glance to Fig.\ref{fig2} shows, when $T_{c}$ vanishes in the
underdoped limit, the anisotropy $\gamma _{T}$ tends to infinity.
Accordingly a 2D-QSI transition is expected to occur. In the
overdoped limit, $T_{c}$ vanishes again but the finite anisotropy
implies a 3D-QSN transition. Since the aforementioned empirical
correlations turned out to be remarkably generic (see
Fig.\ref{fig2}) they appear to reflect universal properties
characterizing these quantum phase transitions. Indeed, the
empirical correlation (\ref{eq1a}) points with the scaling law
(\ref{eq14}) to $\ z\overline{\nu }=1$ in both transitions.
Moreover, the empirical relation between $T_{c}$ and $\gamma _{T}$
(\ref {eq1c}) implies according to the scaling law (\ref{eq22}) a
2D-QSI transition with $\overline{\nu }=1$. Thus, the universality
classes emerging from the empirical relations are characterized by
the critical exponents:
\begin{equation}
\text{2D-QSI}:z=1,\ \overline{\nu }=1,  \label{eq39a}
\end{equation}
\begin{equation}
\text{3D-QSN}:z\overline{\nu }=1.  \label{eq39b}
\end{equation}
These 2D-QSI exponents are consistent with the theoretical
prediction for a 2D disordered bosonic system with long-range
Coulomb interaction. Here the loss of superfluidity is due to the
localization of the pairs, which is ultimately responsible for the
transition\cite{fisher,herbut}. A potential candidate for the
3D-QSN transition is the Ginzburg-Landau theory proposed by Herbut
\cite{herbutd}. It describes a disordered d-wave superconductor to
metal transition at weak coupling and is characterized by the
critical exponents $z=2$ and $\overline{\nu }=1/2$, except in an
exponentially narrow region. Since the resulting $z\overline{\nu
}$ coincides with the value implied by the empirical correlation
(\ref{eq1a}), one expects with the estimate (\ref{eq39b}),
\begin{equation}
\text{3D-QSN}:z=2,\ \overline{\nu }=1/2.  \label{eq39c}
\end{equation}

\begin{figure}[tbp]
\centering
\includegraphics[totalheight=7cm]{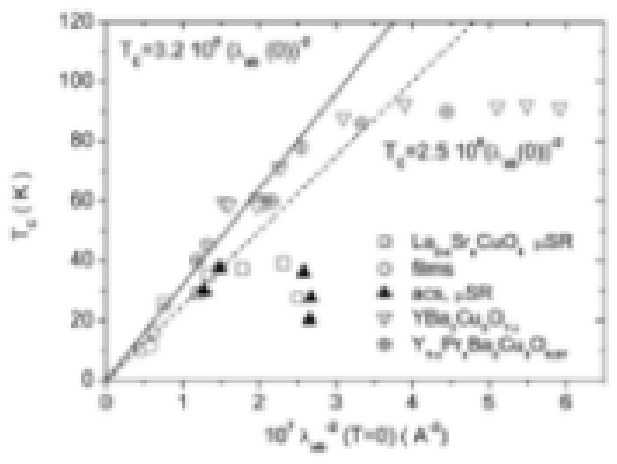}
\caption{$T_{c}$ versus $\protect\lambda _{ab}\left( T=0\right)
^{-2}$ for La $_{2-x}$Sr$_{x}$CuO$_{4}$ ( $\bigcirc $:
\protect\cite{perret}, $\blacktriangle $:
\protect\cite{panagopoulos}, $\Box $: \protect\cite{uemura214})\
YBa$_{2}$Cu$_{3}$O$_{7-\protect\delta }$ ( $\bigtriangledown $:
\protect\cite{zimmermann}) and
Y$_{1-x}$Pr$_{x}$Ba$_{2}$Cu$_{3}$O$_{6.97}$ ($\otimes $:
\protect\cite{seaman}). The dashed and solid lines correspond to
Eq.\ref{eq39}.}
\label{fig14}
\end{figure}

A characteristic property of a 2D-QSI transition, irrespective of
the value of $z\overline{\nu }$, is the universal relation
(\ref{eq16}) between transition temperature and zero temperature
penetration depth. An instructive example is the onset of
superfluidity in $^{4}$He films adsorbed on disordered substrates,
where the linear relationship between $T_{c}$ and
aerial superfluid density $n_{s}^{\square }\left( 0\right) \propto d_{s}/{%
\lambda _{ab}^{2}}\left( {0}\right) $ is well
confirmed\cite{crowell}. In cuprates, a nearly linear relationship
between $T_{c}$ and $\lambda _{ab}\left( T=0\right) ^{-2}$ has
been established some time ago by Uemura {\em et al.}
\cite{uemura214}. In the present context, it is not strictly
universal, because $d_{s}$, the thickness of the independent
slabs, is known to adopt family dependent
values\cite{book,schneisingcross}.\ This fact can be anticipated
from Fig.\ref{fig14}, showing experimental data for $T_{c}$
versus $1/\lambda _{ab}^{2}\left( T=0\right) $ of \ La$_{2-x}$Sr$_{x}$CuO$%
_{4}$ \cite{panagopoulos,uemura214,perret},
YBa$_{2}$Cu$_{3}$O$_{7-\delta }$ \cite{zimmermann}\ and
Y$_{1-x}$Pr$_{x}$Ba$_{2}$Cu$_{3}$O$_{6.97}$ \cite {seaman}. With
$T_{c}$ in K and $\lambda _{ab}(T=0)$ in A, the dashed and solid
straight lines correspond to
\begin{equation}
T_{c}\approx \frac{3.2\ 10^{8}}{\lambda _{ab}^{2}\left( T=0\right) },\ \frac{%
2.5\ 10^{8}}{\lambda _{ab}^{2}\left( T=0\right) }.  \label{eq39}
\end{equation}
Invoking then the universal relation (\ref{eq16}) we obtain the
estimate,
\begin{equation}
\frac{d_{s}\left( \text{YBa}_{2}\text{Cu}_{3}\text{O}_{7-\delta }\right) }{%
d_{s}\left( \text{La}_{2-x}\text{Sr}_{x}\text{CuO}_{4}\right) }\approx \frac{%
3.2}{2.5}\approx 1.3,  \label{eq40}
\end{equation}
which is consistent with , $d_{s}\left( \text{YBa}_{2}\text{Cu}_{3}\text{O}%
_{7-\delta }\right) \ /\ d_{s}\left( \text{La}_{2-x}\text{Sr}_{x}\text{CuO}%
_{4}\right) \approx 10.1\ A/7.6\ A\approx 1.33$, derived from the
thickness tuned QSI transition and the crossing point phenomenon,
respectively \cite {book,schneisingcross}. Consequently,
$dT_{c}/d(1/\lambda _{\Vert }^{2}(T=0)) $ is not strictly
universal. Nevertheless, due to the small variations of $%
d_{s}$ within a family of
cuprates it adopts there a nearly unique value. For this reason the empirical proportionality of $%
T_{c}$ and $\lambda _{ab}\left( T=0\right) ^{-2}$ for underdoped
members of a given family confirms the flow to 2D-QSI criticality.

\begin{figure}[tbp]
\centering
\includegraphics[totalheight=5cm]{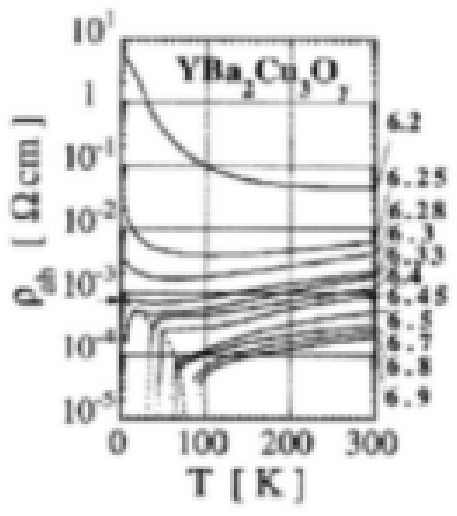}
\includegraphics[totalheight=5cm]{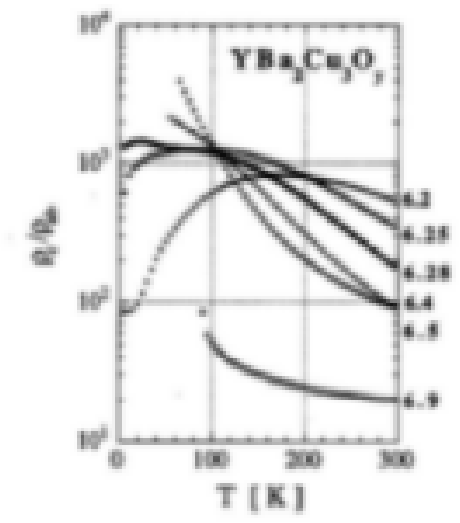}
\caption{Left panel: Temperature dependence of the in-plane
resistivity $\protect\rho_{ab}$ of YBa$_{2}$Cu$_{3}$O$_{y}$ at
various dopant concentrations $y$. Taken from Semba and
Matsuda\protect\cite{semba}. The threshold resistivity is
indicated by an arrow. Right panel: $\protect\rho
_{c}/\protect\rho _{ab}$ versus $T$ of underdoped
YBa$_{2}$Cu$_{3}$O$_{y}$. Taken from Semba and Matsuda
\protect\cite{semba}.} \label{fig15}
\end{figure}

Next we turn to the behavior of the resistivity close to 2D-QSI
criticality. In Fig.\ref{fig15} we displayed the data of Semba and
Matsuda \cite{semba} for the temperature dependence of the
in-plane resistivity $\rho _{ab}$ of YBa$ _{2}$Cu$_{3}$O$_{y}$ at
various dopant concentrations $y$. Below $y\approx 6.3$ the
resistivity increases with decreasing temperature, signaling the
onset of insulating behavior in the zero temperature limit. Above
$y\approx 6.3$ and as the temperature is reduced, the resistivity
drops rapidly and vanishes at and below $T_{c}$. Thus, for
$y\gtrsim 6.3$ there is a superconducting phase and the
2D-QSI-transition occurs at $y\approx 6.3$. Moreover, the
temperature dependence of $\rho _{c}/\rho _{ab}$, depicted in
Fig.\ref{fig15}, is in accord with the scaling relations
(\ref{eq14}) and (\ref{eq28}), yielding $\rho _{c}/\rho
_{ab}=\gamma _{T_{c}}^{2}\propto \delta ^{-2\overline{\nu
}}\propto T_{c}^{2/z}$. Indeed, the anisotropy increases by
approaching the 2D-QSI transition. In contrast, near the 2D-QSI
transition ($y\approx 6.3$), there is a finite threshold in-plane
resistivity $\rho _{ab}^{th}$ $\approx 0.8m\Omega cm$ (see
Fig.\ref{fig15}). According to the scaling relation (\ref{eq30})
this is a characteristic feature of a 2D-QSI transition. For
$d_{s}\approx 11.8A$ and $\sigma _{0}\approx 1$ this leads to the
sheet resistance $\rho _{ab}^{th}/d_{s}\approx h/\left(
4e^{2}\right) \approx 6.5\ \ k\Omega $. Since $\rho _{c}\propto
\gamma ^{2}\rho _{ab}$ and $\rho _{ab}^{th}\rightarrow
d_{s}h/\left( 4e^{2}\right) $ it also becomes evident that the
rise of $\rho _{c}$ for $T>T_{c}$ simply reflects the increasing
anisotropy and with that the flow to 2D-QSI criticality. Together
with the doping dependence of $\gamma _{Tc}$ (see Figs.\ref{fig1}
and \ref{fig2}), these features clearly confirm the occurrence of
a 2D-QSI transition.
\begin{figure}[tbp]
\centering
\includegraphics[totalheight=5cm]{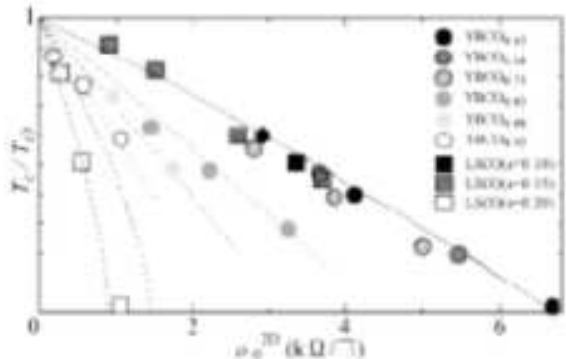}
\caption{Residual in-plane resistance versus normalized critical
temperature for Zn-substituted La$_{2-x}$Sr$_{x}$CuO$_{4}$ and
YBa$_{2}$Cu$_{3}$O$_{7-\protect\delta }$. $T_{c0}$ is the
transition temperature of the Zn free compound. Taken from
Fukuzumi {\em et al.} \protect\cite{fukuzumi}.}
\label{fig16}
\end{figure}

According to Eqs.(\ref{eq16}) and (\ref{eq38})a 2D-QSI-transition
is also characterized by the scaling relation
\begin{equation}
k_{B}T_{c}=c\left( \frac{\rho _{0c}-\rho _{0}}{\rho _{0c}}\right)
^{z\ \overline{\nu }}={\frac{\Phi _{0}^{2}}{16\pi
^{3}\overline{Q}_{2}}}{\frac{d_{s}}{\lambda _{ab}^{2}(T=0)},}
\label{eq41}
\end{equation}
where $\rho _{0}$ and $\rho _{0c}$ denote the residual and
critical residual sheet resistivity, respectively. This prediction
is well confirmed by the data of Fukuzumi {\em et al.} for
Zn-substituted La$_{2-x}$Sr$_{x}$CuO$_{4} $ and
YBa$_{2}$Cu$_{3}$O$_{7-\delta }$ \cite{fukuzumi} displayed in
Fig.\ref{fig16}. Approaching the underdoped limit the data merge
on a straight line. With the scaling relation (\ref{eq41}) this
points to $z\overline{\nu }=1$ consistent with the value emerging
from the empirical correlations (Eq.(\ref{eq39a})).

\begin{figure}[tbp]
\centering
\includegraphics[totalheight=5cm]{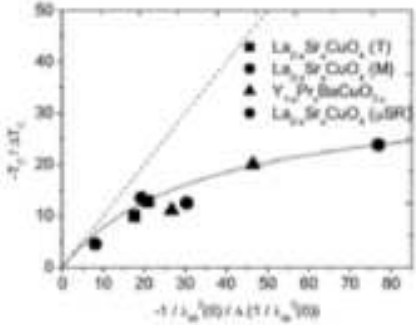}
\caption{Oxygen isotope effect for underdoped
La$_{2-x}$Sr$_{x}$CuO$_{4}$ in terms of $\ -T_{c}/\Delta T_{c}$ \
versus $\ -\left( 1/{\protect\lambda _{ab}^{2}(0)}\right) /\Delta
\left( 1/{\protect\lambda _{ab}^{2}(0)}\right) $. The solid line
marks the critical behavior at the 2D-QSI transition
(Eq.(\ref{eq43})), while the dashed curves indicates the crossover
to criticality. Experimental data taken from: $\bullet $
\protect\cite{zhao}, $ \blacksquare $ \protect\cite{hoferprl} and
$\blacktriangle $ \protect\cite {khasanov}.}
\label{fig17}
\end{figure}

\begin{figure}[tbp]
\centering
\includegraphics[totalheight=5cm]{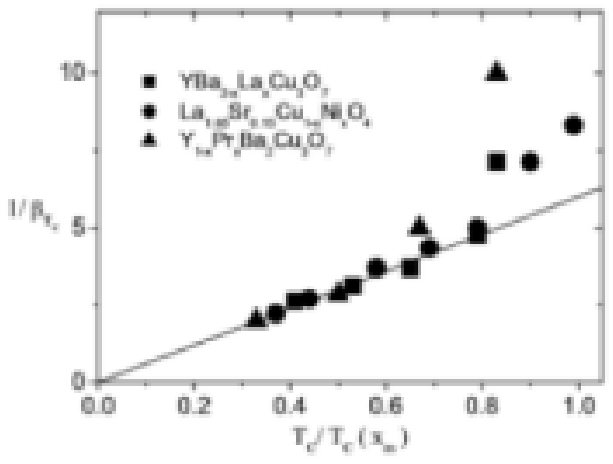}
\caption{Inverse isotope coefficient $1/\protect\beta _{T_{c}}$
versus $ T_{c}/T_{c}^{m}$ for
Y$_{1-x}\Pr_{x}$Ba$_{2}$Cu$_{3}$O$_{7}$ \protect\cite {franck},
La$_{1.85}$Sr$_{0.15}$Cu$_{1-x}$Ni$_{x}$O$_{4}$ \protect\cite
{babushkina} and YBa$_{2-x}$La$_{x}$Cu$_{3}$O$_{7}$
\protect\cite{bornemann}. The straight line corresponds to
$1/\protect\beta _{T_{c}}=$ $\overline{r} \ T_{c}/T_{c}^{m}$
(Eq.(\ref{eq46})) with $z\overline{\protect\nu }=1$ and $
\overline{r}=6$.} \label{fig18}
\end{figure}

Next we turn to the isotope effect. Since universal relations like
(\ref{eq4}), (\ref{eq8}) and (\ref{eq16}) should apply
irrespective of the doping and substitution level, the isotope
effects on the quantities involved, are not independent. As an
example we consider the universal relation (\ref{eq16}),
predicting that close to the 2D-QSI transition the isotope effect
on transition temperature and zero temperature in-plane
penetration depth are related by
\begin{equation}
\frac{\Delta T_{c}}{T_{c}}=\frac{\Delta \left( 1/\lambda
_{ab}^{2}\left( 0\right) \right) }{1/\lambda _{ab}^{2}\left(
0\right) },  \label{eq43}
\end{equation}
where $\Delta \left( B\right) $ denotes the shift of $B$ upon
isotope substitution. Although the available experimental data on
identical samples are rather sparse, the results shown in
Fig.\ref{fig17} for the oxygen isotope effect in
La$_{2-x}$Sr$_{x}$Cu$_{1-x}$O$_{4}$ \cite{zhao,hoferprl} and
Y$_{1-x}\Pr_{x}$Ba$_{2}$Cu$_{3}$O$_{7}$ clearly reveal the
crossover to the asymptotic 2D-QSI behavior marked by the straight
line. An essential result is that the flow to 2D-QSI criticality
implies that the isotope coefficients
\begin{equation}
\beta _{T_{c}}=-\frac{m}{T_{c}}\frac{\Delta T_{c}}{\Delta m},\ \
\beta _{\lambda }=-\frac{m}{T_{c}}\frac{\Delta \left( 1/\lambda
_{ab}^{2}\left( 0\right) \right) }{\Delta m},  \label{eq44}
\end{equation}
diverge. Some insight is obtained by noting that in the doping
regime of interest, isotope substitution does not affect the
dopant and substitution concentrations. \cite{zhao}. In contrast
it lowers the transition temperature and shifts the underdoped
limit $x_{u}$\cite{tskel,franck}. From the relation $T_{c}=a\
\delta ^{z\overline{\nu }}$ (Eq.(\ref{eq14})), yielding
\begin{equation}
\frac{\Delta T_{c}}{T_{c}}=-\frac{z\overline{\nu
}}{x/x_{u}-1}\frac{\Delta x_{u}}{x_{u}},  \label{eq45}
\end{equation}
we obtain for the isotope coefficient the expression
\begin{equation}
\beta _{T_{c}}=\frac{1}{\overline{r}}\left( \frac{T_{c}\left(
x_{m}\right) }{T_{c}}\right) ^{1/z\overline{\nu }},\ \
\frac{1}{\overline{r}}=z\overline{\nu }\frac{m}{\Delta
m}\frac{\Delta x_{u}}{x_{u}}\left( \frac{a}{T_{c}\left(
x_{m}\right) }\right) ^{1/z\overline{\nu }}, \label{eq46}
\end{equation}
applicable close to the 2D-QSI transition. Here we rescaled
$T_{c}$ by $T_{c}\left( x_{m}\right) $, the transition temperature
at optimum doping, to reduce variations of $T_{c}$ between
different materials \cite{tskel}. In Fig.\ref{fig18} we show the
experimental data for Y$_{1-x}\Pr_{x}$Ba$_{2}$Cu$_{3}$O$_{7}$
\cite{franck}, La$_{1.85}$Sr$_{0.15}$Cu$_{1-x}$Ni$_{x}$O$_{4}$
\cite{babushkina} and YBa$_{2-x}$La$_{x}$Cu$_{3}$O$_{7}$
\cite{bornemann} in terms of $1/\beta _{T_{c}}$ versus
$T_{c}/T\left( x_{m}\right) $. As predicted by Eq.(\ref{eq46}),
approaching the 2D-QSI transition $ T_{c}/T_{c}\left( x_{m}\right)
=0$, the data collapse on a straight line, pointing again to
$z\overline{\nu }\approx 1$(Eq.(\ref{eq39a})). Accordingly, the
strong doping dependence of the isotope coefficients of transition
temperature and zero temperature in-plane penetration depth in
underdoped cuprates follows naturally from the doping tuned 3D-2D
crossover and the associated 2D-QSI transition in the underdoped
limit. One might hope that this novel point of view about the
isotope effects in cuprate superconductors \cite{schneikell} will
stimulate further experimental work to obtain new data to confirm
or refute these predictions.

\begin{figure}[tbp]
\centering
\includegraphics[totalheight=3.5cm]{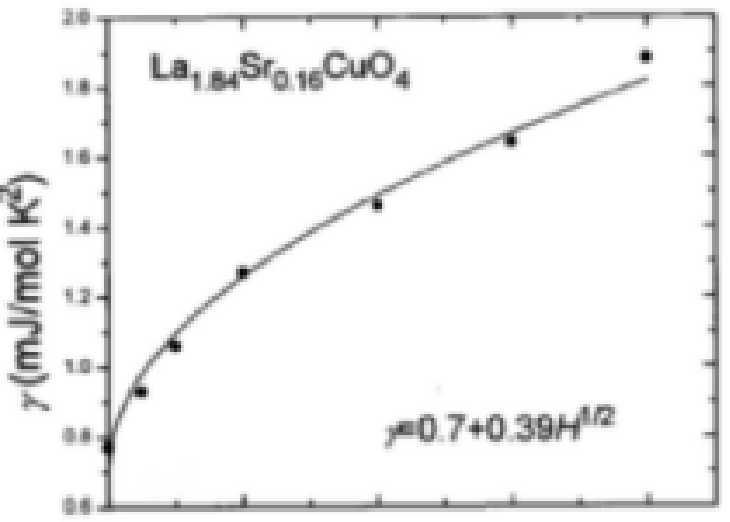}
\includegraphics[totalheight=3.5cm]{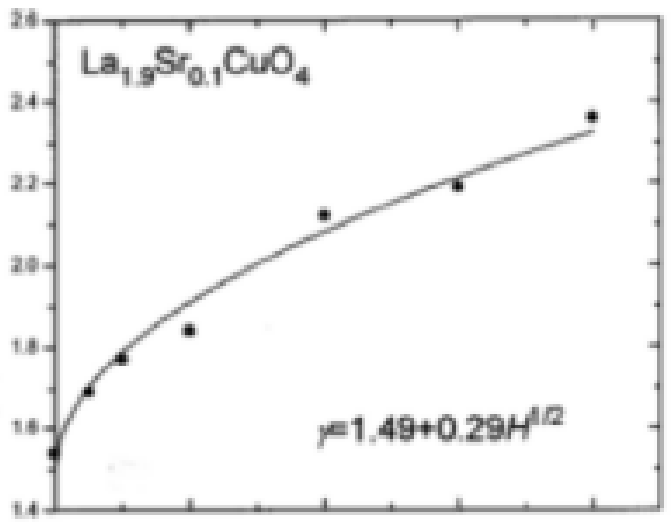}
\includegraphics[totalheight=3.5cm]{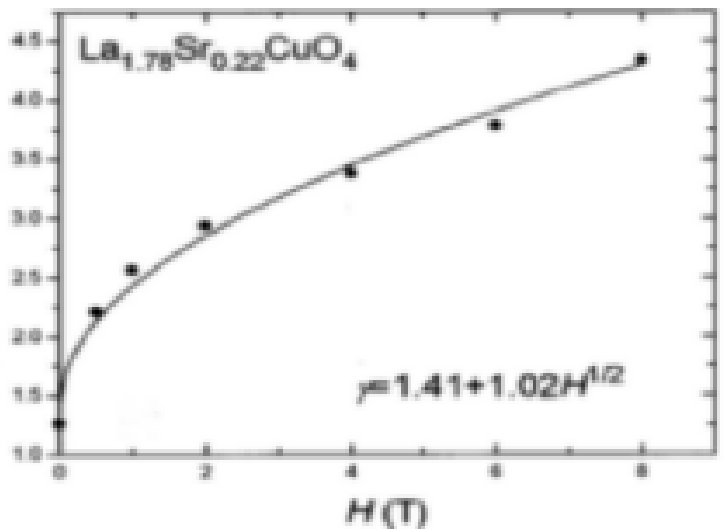}
\caption{Magnetic field dependence of the specific heat
coefficient at zero temperature $\left. \protect\gamma _{c}\right|
_{T=0}$ for La$_{2-x}$Sr$_{x}$CuO$_{4}$ at x=0.1, 0.16 and 0.22.
Taken from Chen {\em et al.}\protect\cite {chen}.}
\label{fig19}
\end{figure}

A property suited to shed light on the critical behavior of both,
the 2D-QSI and 3D-QSN transition is the magnetic field dependence
of the specific heat coefficient in the limit of \ zero
temperature. From the scaling relation $\left. \gamma _{c}\right|
_{T=0}\propto H_{c}^{\left( D-z\right) /2}$ (Eq.(\ref{eq20})) it
is seen that for both transition $\left. \gamma _{c}\right|
_{T=0}\propto H_{c}^{1/2}$ holds, provided that $z=1$ and $z=2$ at
2D-QSI and 3D-QSN criticality, respectively. The experimental data
displayed in Fig.\ref{fig19} shows that in
La$_{2-x}$Sr$_{x}$CuO$_{4}$ $\left( D-z\right) /2=1/2$ holds
irrespective of the doping level. Thus these data provide rather
unambiguous evidenced for a 2D-QSI transition with $z=1$ and a
3D-QSN criticality with $z=2$. This implies that $\left. \gamma
_{c}\right| _{T=0}\propto H_{c}^{1/2}$ is not a characteristic
feature of d-wave pairing, as proposed by Volovik\cite {volovik},
Won and Maki\cite{won}.

\begin{figure}[tbp]
\centering
\includegraphics[totalheight=5cm]{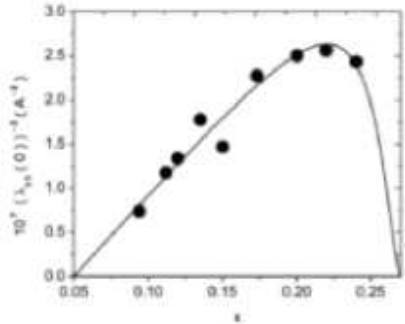}
\caption{$1/\protect\lambda _{ab}^{2}\left( 0\right) $ versus $x$
for La$_{2-x}$Sr$_{x}$CuO$_{4}$. $\bullet $: experimental data
taken from \protect\cite{panagopoulos,uemura214}. The solid curve
is a fit to Eq.(\ref {eq48}) with the parameters listed in
Eq.(\ref{eq49}). It indicates the crossover from a 2D-QSI
transition with $z=1$ and $\overline{\protect\nu }=1$ to a 3D-QSN
criticality with $z=2$ and $\overline{\protect\nu }=1/2$.}
\label{fig20}
\end{figure}

The above comparison with experiment provides rather clear
evidence for the occurrence of a 2D-QSI transition with $z=1$ and
$\overline{\nu }=1$ in the underdoped and a 3D-QSI critical point
$z=2$ and $\overline{\nu }=1/2$ in the overdoped limit. Next, to
substantiate this scenario further, we consider the crossover
between these quantum critical points. Noting that $ 1/\lambda
_{ab}^{2}\left( 0\right) $ scales close to these critical points
as (Eq.(\ref{eq13}))
\begin{equation}
\frac{1}{\lambda _{ab}^{2}\left( 0\right) }\propto \Upsilon
_{ab}^{D}\left( 0\right) \propto \delta ^{\overline{\nu }\left(
\left( D-2+z\right) \right) },  \label{eq47}
\end{equation}
we invoke
\begin{equation}
\lambda _{ab}^{2}\left( 0\right) =\frac{a_{\lambda
}}{x-x_{u}}+\frac{ b_{\lambda }}{\left( x_{o}-x\right) ^{3/2}},
\label{eq48}
\end{equation}
to interpolate between the 2D-QSI and 3D-QSN transition. A fit to
the experimental data of La$_{2-x}$Sr$_{x}$CuO$_{4}$, yielding the
parameters
\begin{equation}
a_{\lambda }=5.42\ 10^{5}\ A^{2},\ \ b_{\lambda }=6.9\ 10^{3}\
A^{2}, \label{eq49}
\end{equation}
is shown Fig.\ref{fig20}.

\begin{figure}[tbp]
\centering
\includegraphics[totalheight=5cm]{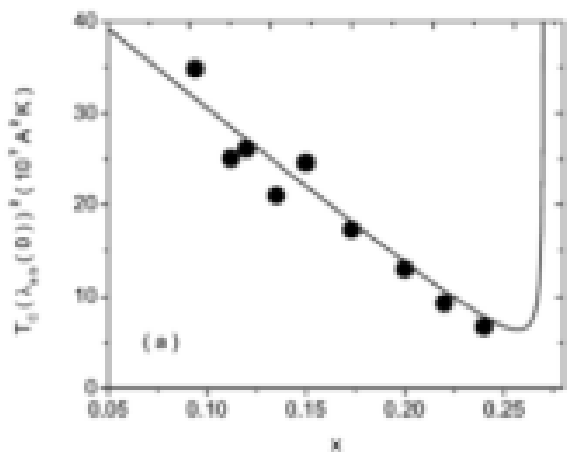}
\includegraphics[totalheight=5cm]{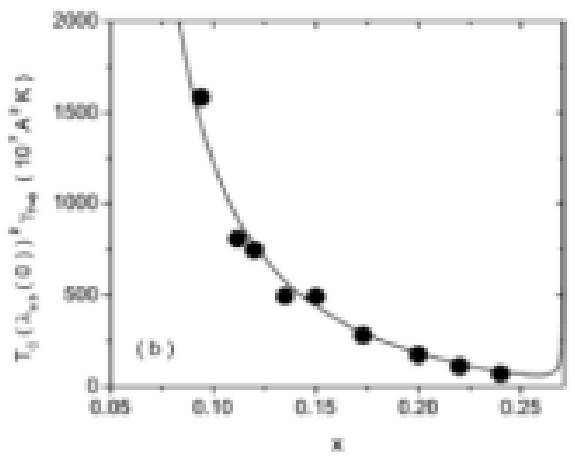}
\caption{(a) $T_{c}\protect\lambda _{ab}^{2}\left( 0\right) $
versus $x$ for for La$_{2-x}$Sr$_{x}$CuO$_{4}$. $\bullet $: taken
from \protect\cite {panagopoulos} and \protect\cite{uemura214}.
The solid line is Eqs.(\ref{eq1a}) and (\ref{eq48}) with
$T_{c}\left( x_{m}\right) =39.8 K$ and the parameters listed in
Eq.(\ref{eq49}). Note that close to the 2D-QSI and 3D-QSN
transition $T_{c}\protect\lambda _{ab}^{2}\left( 0\right) \propto
\overline{\protect\xi }_{c}$. (b) $T_{c}\protect\lambda
_{ab}^{-2}\left( 0\right) \protect\gamma _{T=0} $ versus $x$ for
La$_{2-x}$Sr$_{x}$CuO$_{4}$. $\bullet $\ : $T_{c}$ and
$\protect\lambda _{ab}^{-2}\left( 0\right) $ taken from
\protect\cite {panagopoulos,uemura214} and $\protect\gamma _{T=0}$
from Eq.(\ref{eq1b}) with $ \protect\gamma _{T=0,0}=2$. Note that
$T_{c}\protect\lambda _{ab}^{-2}\left( 0\right) \protect\gamma
_{T=0}\propto \overline{\protect\xi }_{ab}$(\ref {eq24}). The
solid curve indicates the crossover from 2D-QSI $\left( z=1,\
\overline{\protect\nu }=1\right) $ to 3D-QSN $\left( z=2,\
\overline{\protect \nu }=1/2\right) $ criticality according to
Eqs.(\ref{eq1a}), (\ref{eq1b}), (\ref{eq47}) and (\ref{eq48}).}
\label{fig21}
\end{figure}

\begin{figure}[tbp]
\centering
\includegraphics[totalheight=5cm]{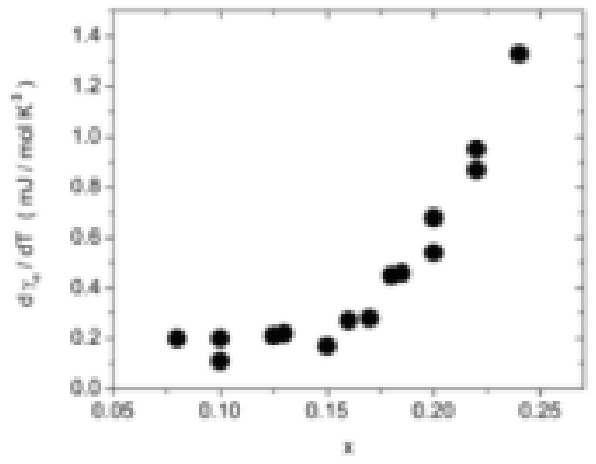}
\caption{$\left. T_{c}d\protect\gamma _{c}/dT\right| _{T=0}$
versus $x$ for La$_{2-x}$Sr$_{x}$CuO$_{4}$.Data taken from
\protect\cite{momono} and \protect\cite{loram}.} \label{fig22}
\end{figure}

It is remarkable that this simple interpolation scheme, reducing
in the under- and overdoped limit to the expected asymptotic
behavior, describes the data so well. Due to this agreement, this
interpolation function provides in conjunction with the empirical
law for $ T_{c}\left( x\right) $ (Eq.(\ref{eq1a})), a realistic
description of the doping dependence of the zero temperature
out-of-plane correlation length $\overline{\xi }_{c}$, given by
Eq.(\ref{eq24}), yielding $\overline{\xi } _{c}^{-}\propto
T_{c}\lambda _{ab}^{2}\left( 0\right) $, close to 2D-QSI and
3D-QSN criticality. The doping dependence of $T_{c}\lambda
_{ab}^{2}\left( 0\right) $ is displayed in Fig.\ref{fig21} for
La$_{2-x}$Sr$_{x}$CuO$_{4}$. Since $T_{c}\propto d_{s}/\lambda
_{ab}^{2}\left( 0\right) $ holds in the underdoped limit (see
Eq.(\ref{eq16}) and Fig.\ref{fig14}), it is clear that $
T_{c}\lambda _{ab}^{2}\left( 0\right) $ tends to a constant value,
proportional to $d_{s}$, the thickness of the independent sheets.
Since initially $d/dx\left( T_{c}\lambda _{ab}^{2}\left( 0\right)
\right) \approx -T_{c}\left( x_{m}\right) a_{\lambda }x_{o}$, the
linear decrease simply reflects the parabolic form of the
empirical law (\ref{eq1a}). Finally, the upturn close to the
overdoped limit, a regime which experimentally has not yet been
attained, signals the 3D-QSN transition, where $\overline{\xi }
_{c}\propto T_{c}\lambda _{ab}^{2}\left( 0\right) $ diverges. In
this context it is also instructive to consider the zero
temperature in-plane correlation length. By definition it diverges
at a 2D and 3D quantum phase transition. According to the scaling
relation (\ref{eq24}) it can be measured in terms of
$\overline{\xi }_{ab}\propto T_{c}\lambda _{ab}^{2}\left( 0\right)
\gamma _{T=0}$. In Fig.\ref{fig21} we displayed the experimental
data for $\overline{\xi }_{ab}\propto T_{c}\lambda _{ab}^{2}\left(
0\right) \gamma _{T=0}$ versus $x$. For comparison we included the
behavior resulting from Eqs.(\ref{eq1a}), (\ref{eq1b}),(\ref
{eq47}) and (\ref{eq48}) in terms of the solid curve. While the
rise of $ \overline{\xi }_{ab}$ in the underdoped regime,
signaling the occurrence of a 2D-QSI transition, is well
confirmed, it does not extend sufficiently close to the overdoped
limit to detect 3D-QSN criticality. From Fig.\ref {fig22}, showing
the doping dependence of the $T$-linear term of the specific heat
coefficient of La$_{2-x}$Sr$_{x}$CuO$_{4}$, it is seen that this
quantity tends to a finite value in the underdoped limit and
increases in the overdoped regime. This behavior is fully
consistent with the scaling relation $\left. d\gamma
_{c}/dT\right| _{T=0}\propto \delta ^{\overline{\nu }\left(
D-2z\right) }$ (Eq.(\ref{eq21})). Indeed at 2D-QSI criticality
with $ z=1$, $\left. d\gamma _{c}/dT\right| _{T=0}$ tends to a
constant value and diverges close to the 3D-QSN critical point for
$z>3/2$. Unfortunately, the data does not extend sufficiently
close to the overdoped limit to provide reliable estimates of the
exponent combination $\overline{\nu }\left( 3-2z\right) $.

Provided that the linear T-term of $1/\lambda _{ab}^{2}\left(
T\right) $ in the zero temperature limit exists, the scaling
relations (\ref{eq13}) and (\ref{eq14}) yield close to quantum
criticality the universal relation
\begin{equation}
T_{c}\left. \frac{d}{dT}\left( \frac{\lambda _{i}(0)}{\lambda
_{i}(T)}\right) ^{2}\right| _{T=0}=\overline{y}_{c}\left.
\frac{d{Y}_{D}}{d\overline{y}}\right| _{\overline{y}=0},
\label{eq50}
\end{equation}
where D=2 and D=3 for the 2D-QSI and 3D-QSN transition,
respectively. In Table I we collected additional estimates for
various cuprates and doping levels. The rise of $\ $the magnitude
of $T_{c}\left. d/dT\left( \lambda _{ab}(\delta ,0)/\lambda
_{i}(\delta ,T)\right) ^{2}\right| _{T=0}$ with increasing doping
level reflects the 2D-3D-crossover in the scaling function
$Y_{D}$. Noting that most compounds are close to optimum doping it
is not surprising that the listed values scatter around -0.61, the
value of nearly optimally doped La$_{2-x}$Sr$_{x}$CuO$_{4}$.

\begin{center}
\begin{tabular}{|l|l|l|l|}
\hline
& $\lambda _{ab}\left( 0\right) $ (A) & T$_{c}$ (K) & $T_{c}\left. \frac{d}{%
dT}\left( \frac{\lambda _{i}(\delta ,0)}{\lambda _{i}(\delta
,T)}\right) ^{2}\right| _{T=0}$ \\ \hline
HgBa$_{2}$Ca$_{2}$Cu$_{3}$O$_{8+\delta }$ & 1770 & 135 & -0.59 \\
\hline HgBa$_{2}$CuO$_{4+\delta }$ & 1710 & 93 & -0.65 \\ \hline
La$_{1.9}$Sr$_{0.1}$CuO$_{4}$ & 2800 & 30 & -0.49 \\ \hline
La$_{1.85}$Sr$_{0.15}$CuO$_{4}$ & 2600 & 39 & -0.61 \\ \hline
La$_{1.8}$Sr$_{0.2}$CuO$_{4}$ & 1970 & 35 & -0.72 \\ \hline
La$_{1.78}$Sr$_{0.22}$CuO$_{4}$ & 1930 & 27.5 & -0.72 \\ \hline
La$_{1.76}$Sr$_{0.24}$CuO$_{4}$ & 1940 & 19 & -0.94 \\ \hline
Bi$_{2}$Sr$_{2}$CaCu$_{2}$O$_{8+\delta }$ & 2600 & 91 & -0.61 \\
\hline Y$_{0.94}$Ca$_{0.06}$Ba$_{2}$Cu$_{4}$O$_{8}$ & 1361 & 88 &
-0.51 \\ \hline YBa$_{2}$Cu$_{4}$O$_{8}$ & 1383 & 81 & -0.58 \\
\hline YBa$_{1.925}$La$_{0.075}$Cu$_{4}$O$_{8}$ & 1521 & 74 &
-0.61 \\ \hline YBa$_{1.9}$La$_{0.1}$Cu$_{4}$O$_{8}$ & 1593 & 72 &
-0.63 \\ \hline
\end{tabular}

\bigskip
\end{center}

Table I: Estimates for $T_{c}\left. d/dT\left( \lambda _{ab}\left(
0\right) /\lambda _{ab}\left( T\right) \right) ^{2}\right| _{T=0}$
derived from the experimental data of: slightly underdoped
HgBa$_{2}$Ca$_{2}$Cu$_{3}$O$_{8+\delta }$, slightly overdoped
HgBa$_{2}$CuO$_{4+\delta }$ \cite {panagophg}, under-, optimally-
and over-doped La$_{2-x}$Sr$_{x}$CuO$_{4}$ \cite{panagopoulos},
Bi$_{2}$Sr$_{2}$CaCu$_{2}$O$_{8+\delta }$ \cite{jacobs},
Y$_{0.94}$Ca$_{0.6}$Ba$_{2}$Cu$_{4}$O$_{8}$,
YBa$_{2}$Cu$_{4}$O$_{8}$, Y$_{1.925}$La$_{0.075}$Cu$_{4}$O$_{8}$
and Y$_{1.9}$La$_{0.1}$Cu$_{4}$O$_{8}$ \cite{shengelaya}.

\bigskip

\begin{figure}[tbp]
\centering
\includegraphics[totalheight=5cm]{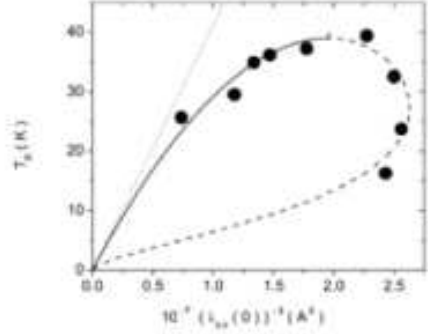}
\caption{$T_{c}$ versus $1/\protect\lambda _{ab}^{2}\left(
0\right) $ for La$_{2-x}$Sr$_{x}$CuO$_{4}$. $\bullet $:
experimental data taken from
\protect\cite{panagopoulos,uemura214}. The solid and dashed curves
result from the empirical law (\ref{eq1a}) and the interpolation
function (\ref{eq48}) with $T_{c}\left( x_{m}\right) $=39.8 K and
the parameters listed in Eq.(\ref {eq49}). The solid line
indicates the flow from optimum doping to 2D-QSI criticality and
the dashed one to the 3D-QSN critical point. } \label{fig23}
\end{figure}

\begin{figure}[tbp]
\centering
\includegraphics[totalheight=5cm]{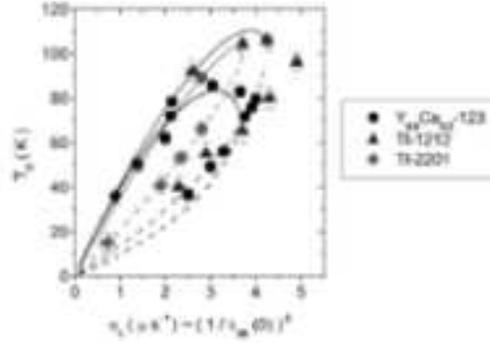}
\caption{$T_{c}$ versus $\protect\sigma _{0}\propto
\protect\lambda _{ab}^{-2}\left( 0\right)$ for
Y$_{0.8}$Ca$_{0.2}$Ba$_{2}$(Cu$_{1-y}$Zn$ _{y}
$)O$_{7-\protect\delta }$ (Y$_{0.8}$Ca$_{0.2}$-123),
Tl$_{0.5-y}$Pb$_{0.5+y}$Sr$_{2}$Ca$_{1-x}$Y$_{x}$Cu$_{2}$O$_{7}$
(Tl-1212)\protect\cite{bernhard} and
TlBa$_{2}$CuO$_{6+\protect\delta }$ (Tl-2201)
\protect\cite{niedermayer}. The solid and dashed curves result
from the empirical law (\ref{eq1a}) and the interpolation function
(\ref {eq48}). The solid curves indicate the flow from optimum
doping to 2D-QSI criticality and the dashed ones to the 3D-QSN
critical point.} \label{fig24}
\end{figure}

Before turning to the substitution tuned quantum transitions it is
useful to express the doping dependence in the interpolation
formula (\ref{eq48}) in terms of the transition temperature. This
is achieved by invoking the empirical correlation \ref{eq1a}
between $T_{c}$ and doping concentration $x$ . In Fig.\ref{fig23}
we displayed the resulting Uemura plot, $T_{c}$ versus $ 1/\lambda
_{ab}^{2}\left( 0\right) $ for La$_{2-x}$Sr$_{x}$CuO$_{4}$ in
terms of the solid and dashed curves, resembling the outline of a
fly's wing. The solid curve marks the flow from $T_{c}\left(
x_{m}\right) $ to the 2D-QSI critical point and the dashed one the
flow to 3D-QSN criticality. The dotted line indicates the
universal 2D-behavior (Eqs.(\ref{eq16})) and (\ref {eq39})). For
comparison we included the experimental data of Panagopoulos {\em
et al.}\cite{panagopoulos} and Uemura {\em et
al.}\cite{uemura214}. Although the data does not attain the
respective critical regimes, together with the theoretical curves,
they provide a generic perspective of the flow to 2D-QSI and
3D-QSN criticality. Indeed convincing evidence for these flows
emerges from the $\mu $SR data displayed in Fig.\ref{fig24} for
Y$_{0.8}$Ca$_{0.2}$Ba$_{2}$(Cu$_{1-y}$Zn$_{y}$)O$_{7-\delta }$
(Y$_{0.8}$Ca$_{0.2}$-123),
Tl$_{0.5-y}$Pb$_{0.5+y}$Sr$_{2}$Ca$_{1-x}$Y$_{x}$Cu$_{2}$O$_{7}$
(Tl-1212) \cite{bernhard} and TlBa$_{2}$CuO$_{6+\delta }\ \left(
\text{Tl-2201}\right) $ \cite{niedermayer}. In analogy to
Fig.\ref{fig23}, the solid curves mark the flow from $T_{c}\left(
x_{m}\right) $ to the 2D-QSI critical point and the dashed ones
the approach to 3D-QSN criticality.

\begin{figure}[tbp]
\centering
\includegraphics[totalheight=5cm]{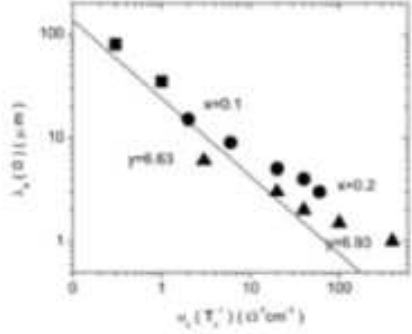}
\caption{$\protect\lambda _{c}\left( 0\right) $ versus
$\protect\sigma_{c}\left( T_{c}^{+}\right) $ for
YBa$_{2}$Cu$_{3}$O$_{7-\protect\delta }$ ($ \blacktriangle $),
La$_{2-x}$Sr$_{x}$CuO$_{4}$ ($\bullet $ )\protect\cite {uchida}
and Bi$_{2}$Sr$_{2}$CaCu$_{2}$O$_{y}$ ($\blacksquare
$)\protect\cite {shibata}. The straight line is Eq.(\ref{eq32})
with $z=1$ and $\Omega _{s}\approx 24\protect\mu m\left( \Omega
cm\right) ^{3/4}$ , the estimate for La$_{2-x}$Sr$_{x}$CuO$_{4}$.}
\label{fig25}
\end{figure}

A generic flow to the 2D-QSI critical point also emerges from the
plot $ \lambda _{c}\left( 0\right) $ versus $\sigma _{c}\left(
T_{c}^{+}\right) $ displayed in Fig.\ref{fig25} for
YBa$_{2}$Cu$_{3}$O$_{7-\delta }$, La$_{2-x}$Sr$_{x}$CuO$_{4}$ and
Bi$_{2}$Sr$_{2}$CaCu$_{2}$O$_{y}$ at various doping levels. An
empirical, nearly linear, correlation between these quantities has
been proposed by Basov {\em et al}\cite{basov}. From the scaling
relation (\ref{eq32}) it is seen that the systematic rise of
$\lambda _{c}\left( 0\right) $ with decreasing $\sigma _{c}\left(
T_{c}^{+}\right) $, tuned by decreasing dopant concentration and
the associated rise of the anisotropy $\gamma _{T}$, reflects
again the flow to 3D-QSI criticality. The straight line marks the
asymptotic behavior La$_{2-x}$Sr$_{x}$CuO$_{4}$ for $z=1$ and
$\Omega _{s}\approx 24\mu m\left( \Omega cm\right) ^{3/4}$. Since
$\Omega _{s}$ depends on the critical amplitudes\ $\gamma _{0,0}$,
$\lambda _{ab,0}\left( 0\right) $, $\gamma _{T_{c},0}$ and the
thickness $d_{s}$ (Eqs.(\ref{eq32}) and (\ref{eq33})), its value
is unique within a family of cuprates. $\Omega _{s}\approx 24\mu
m\left( \Omega cm\right) ^{3/4}$ follows from $\gamma
_{0,0}\approx 1.63$, $\gamma _{T_{c},0}\approx 2$ (see Fig.\ref
{fig1}), $d_{s}\approx 6.6\ A$, $\sigma _{0}\approx 1$ and
$\lambda _{ab,0}\left( 0\right) \approx 736\ A$ (Eq.(\ref{eq49}).
Although the experimental data is still quite far from the
underdoped limit, the flow to 2D-QSI criticality with family
dependent values of $\Omega _{s}$ can be anticipated. This differs
from the mean-field prediction for bulk superconductors in the
dirty limit and layered BCS superconductors, treated as weakly
coupled Josephson junctions (see Eqs.(\ref{eq34}) and
(\ref{eq35})) where $\Omega _{s}\propto T_{c}^{-1/2}$. Moreover,
as the optimally doped and underdoped regimes are approached,
systematic deviations from the straight line behavior appear,
indicating the flow to 3D-QSN-criticality (Eq.(\ref{eq37})).

According to the empirical correlation between $T_{c}$ and
anisotropy $\gamma _{T}$ (see Eq.(\ref{eq1c}) and Fig.\ref{fig2}),
the initial value of the dopant concentration determines whether
the flow leads to 2D-QSI or 3D-QSN criticality. For initially
underdoped cuprates, the rise of $\gamma _{T}$ , tuned by the
doping induced reduction of $T_{c}$, directs the flow to 2D-QSI
criticality. Conversely, in initially overdoped cuprates, the fall
of $\gamma _{T}$ \ to a finite value drives the flow to the 3D-QSN
critical point. As the nature of the quantum phase transitions is
concerned we have seen that the 2D-QSI transition has a rather
wide and experimentally accessible critical region. For this
reason we observed considerable and consistent evidence that it
falls into the same universality class as the onset of
superfluidity in $^{4}$He films in disordered media, corrected for
the long-rangeness of the Coulomb interaction. The resulting
critical exponents, $z=1$ and $\overline{\nu }\approx 1$, are also
consistent with the empirical relations (Eqs.(\ref{eq1a}),
(\ref{eq1b}) and (\ref{eq1c})) and the observed temperature and
magnetic field dependence of the specific heat coefficient in the
limit of zero temperature. These properties also point to a 3D-QSN
transition with $z=2$ and $\overline{\nu }\approx 1/2$, describing
a d-wave superconductor to disordered metal transition at weak
coupling. Here the disorder destroys superconductivity, while at
the 2D-QSI transition it localizes the pairs and with that
destroys superfluidity. Due to the existence of the 2D-QSI and
3D-QSN critical points, the detection of finite temperature 3D-XY
critical behavior will be hampered by the associated crossovers
which reduce the temperature regime where thermal 3D-XY
fluctuations dominate. In any case, our analysis clearly revealed
that the universality of the empirical correlations reflect the
flow to 2D-QSI and 3D-QSN criticality. Moreover, the doping tuned
superconductivity in bulk cuprate superconductors turned out to be
a genuine 3D phenomenon, where the interplay of anisotropy and
superconductivity destroys the latter in the 2D limit.

\subsection{Evidence for substitution tuned quantum phase transitions}

It is well established, both experimentally and theoretically,
that in conventional superconductors (e.g., A15 compounds or the
Chevrel phases) the presence of magnetic impurities depresses the
superconducting transition temperature more efficiently than does
the introduction of nonmagnetic ions \cite{abrikosov}, and this
has been ascribed to the breaking of pairs by the magnetic
impurities. To determine wether the cuprates behave similarly with
respect to magnetic impurities, extensive studies involving the
substitution for Cu by other 3d metals have been performed
\cite{tarascon,xiao}. A result common to all these studies is that
$\ T_{c}$ is depressed in the same manner, independent of wether
the substituent is magnetic or nonmagnetic, and in contrast to
that observed in conventional superconductors.

\begin{figure}[tbp]
\centering
\includegraphics[totalheight=5cm]{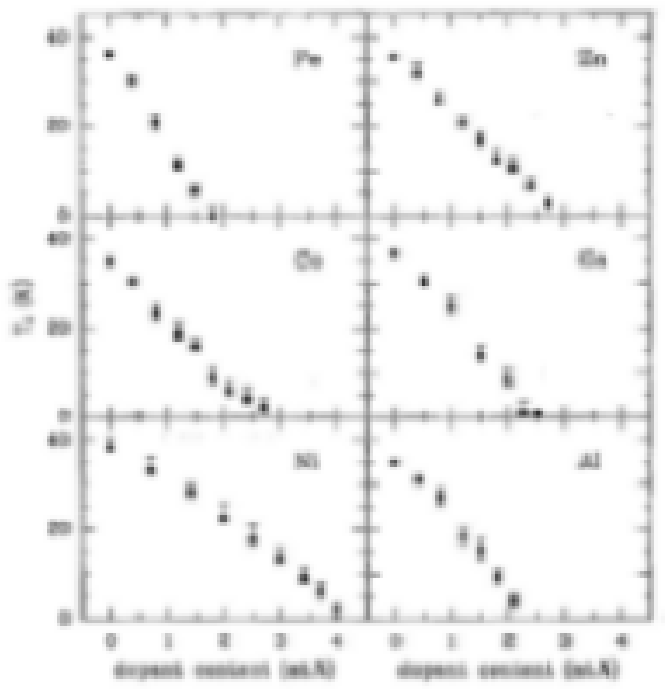}
\caption{Variation of $T_{c}$ with substitution for
La$_{1.95}$Sr$_{0.15}$Cu$_{1-y}$A$_{y}$O$_{4}$ ( A=Fe, Cu, Ni, Zn,
Ga and Al ). Taken from Xiao {\em et al.} \protect\cite{xiao}}
\label{fig26}
\end{figure}

To illustrate this point we depicted in Fig.\ref{fig26}, the
variation of $T_{c} $ with substitution for
La$_{1.95}$Sr$_{0.15}$Cu$_{1-y}$A$_{y}$O$_{4}$ (A=Fe, Cu, Ni, Zn,
Ga and Al) \cite{xiao}. $T_{c}$ is seen to vanish at a critical
substitution concentration $y_{c}$, where a quantum phase
transition occurs. The experimental data for
La$_{2-x}$Sr$_{x}$Cu$_{1-y}$Zn$_{y}$O$_{4}$,\
Y$_{0.8}$Ca$_{0.2}$Ba$_{2}$Cu$_{3-3y}$Zn$_{3y}$O$_{7-\delta }$
\cite{tallonzn} and
Bi$_{2}$Sr$_{2}$CaCu$_{2-2y}$Co$_{2y}$O$_{8+\delta }$
\cite{tallonznbi} shows that this behavior is not restricted to
optimum doping. Thus $T_{c}$ dependence on both, the dopant ($x$)
and substitution concentration ($y$). Fig.\ref{fig3} shows the
resulting ($T_{c},x,y$) diagram for
La$_{2-x}$Sr$_{x}$Cu$_{1-y}$Zn$_{y}$O$_{4}$. The blue curve,
$y_{c}\left( x\right) $, is a line of quantum phase transitions.
At the corresponding critical endpoints the system undergoes a
2D-QSI and 3D-QSN transition. Along the path marked by the green
arrow, a QSI and QSN transition is expected to occur. The pink
arrow indicates the crossover from insulating to metallic
behavior.

\begin{figure}[tbp]
\centering
\includegraphics[totalheight=4cm]{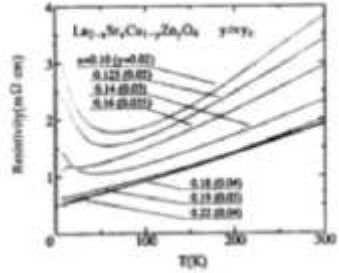}
\caption{Temperature dependence of the resistivity for
La$_{2-x}$Sr$_{x}$Cu$_{1-y}$Zn$_{y}$O$_{4}$ at $y_{c}\left(
x\right) $. Taken from Momono {\em et al.}  \protect\cite
{momono}.}
\label{fig27}
\end{figure}

\begin{figure}[tbp]
\centering
\includegraphics[totalheight=5cm]{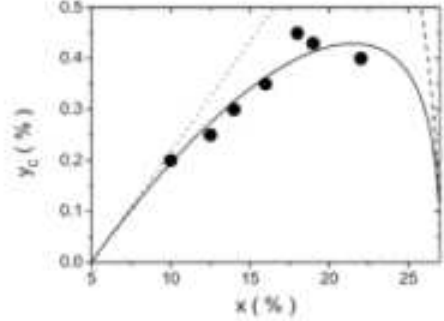}
\caption{$y_{c}$ versus $x$ for
La$_{2-x}$Sr$_{x}$Cu$_{1-y}$Zn$_{y}$O$_{4}$. $\bullet $ : Taken
from Momono {\em et al.} \protect\cite{momono}. The solid curve
interpolates between the critical behavior of a 2D-QSI transition
with $\overline{\protect\nu }=1$ (dotted line: $y_{c}\left(
x\right) =0.03\left( x-5\right) $) and a 3D-QNS transition with
$\overline{\protect\nu }=1/2$ (dashed line: $ y_{c}\left( x\right)
=0.46\left( 27-x\right) ^{1/2}$).} \label{fig28}
\end{figure}

This scenario, emerging from the temperature
dependence of the resistivity of La$_{2-x}$Sr$%
_{x}$Cu$_{1-y}$Zn$_{y}$O$_{4}$\cite{momono}, is shown in
Fig.\ref{fig27}. Along the phase transition line $y_{c}\left(
x\right) $ superconductivity disappears and for $x\gtrsim 0.16$ \
metallic behavior sets in, while for $x\lesssim 0.16$ the
resistivity exhibits insulating behavior at low temperatures. For
fixed $x$ and close to $y_{c}\left( x\right) $, $T_{c}$ scales
according to Eqs.(\ref {eq9}) \ and (\ref{eq14}) as
\begin{equation}
T_{c}\left( y\right) \propto \ \left( 1-\frac{y}{y_{c}\left(
x\right)} \right) ^{z\overline{\nu }}.
\label{eq51}
\end{equation}
Since $z\overline{\nu }=1$ is expected to hold in both, the doping
tuned 2D-QSI and 3D-QSN transition (Eqs.(\ref{eq39a}) and
(\ref{eq39c})), this combination should hold along the entire line
$y_{c}\left( x\right) $. Although the data shown in
Fig.\ref{fig26} is remarkably consistent with $z\overline{\nu }=1$
it remains to be understood, why the linear relationship applies
almost over the entire substitution regime. Close to the 2D-QSI
and 3D-QSN transitions, the singular part of the free energy
density scales in analogy to Eq.(\ref{eq12}) as
\begin{equation}
f_{s}\propto \delta ^{\overline{\nu }\left( D+z\right) }{\cal
F}\left( y\delta ^{-\overline{\nu }}\right) ,\ \delta =x/x_{u}-1,\
1-x/x_{o}. \label{eq52}
\end{equation}
${\cal F}$ is a scaling function of its argument. A phase
transition is signaled by a singularity of the scaling function at
some value of its argument. Thus,
\begin{equation}
y_{c}\left( x\right) \propto \delta ^{\overline{\nu }},
\label{eq53}
\end{equation}
with $\overline{\nu }=1$ close to the 2D-QSI (Eqs.(\ref{eq39a}))
and $\overline{\nu }=1/2$ near the 3D-QSN
transition(Eq.(\ref{eq39c})). The resulting phase transition line
is shown in Fig.\ref{fig28} for
La$_{2-x}$Sr$_{x}$Cu$_{1-y}$Zn$_{y}$O$_{4}$, where we included the
experimental data for comparison. As expected from the doping
dependence of the correlation lengths (see Fig.\ref{fig21})
$y_{c}\left( x\right) $ exhibits close to the 3D-QSN transition a
very narrow critical regime. For this reason the data considered
here is insufficient to provide an estimate for $\overline{\nu }$
. Nevertheless, $\overline{\nu }=1/2$ yields a reasonable
qualitative description of the quantum critical line $y_{c}\left(
x\right) $.

The picture we no have is summarized in the phase diagram shown in
Fig.\ref {fig3}. The blue curve corresponds to the line of quantum
phase transitions $y_{c}\left( x\right) $ shown in
Fig.\ref{fig28}. Along this line, $z\overline{\nu }=1$ is expected
to hold (Eqs.(\ref{eq39a}) and (\ref{eq39c})), while
$\overline{\nu }\approx 1$(Eqs.(\ref{eq39a})) at the 2D-QSI ($y=0$
and $x=x_{u}$) and $\overline{\nu }=1/2$ at the 3D-QSN (($y=0$ and
$x=x_{o}$) transition. According to the empirical correlation
between $T_{c}$ and anisotropy $\gamma _{T}$ (see Eq.(\ref{eq1c})
and Fig.\ref{fig2}), the initial value of the dopant concentration
determines whether the flow upon substitution leads to 2D-QSI or
3D-QSN criticality. For initially underdoped cuprates, the rise of
$\gamma _{T}$ , tuned by the substitution induced reduction of
$T_{c}$, drives the flow to 2D-QSI criticality. Conversely, in
initially overdoped cuprates, the fall of $\gamma _{T}$ \ to a
finite value directs the flow to 3D-QSN criticality. The mechanism
whereby the substitution of Cu leads to a reduction of $T_{c}$ and
finally to the quantum critical line $y_{c}\left( x\right) $
appears to be not well understood. When the aforementioned
universality classes of the 2D-QSI and 3D-QSN transitions hold
true (Eqs.(\ref{eq39a}) and (\ref{eq39c})), disorder plays an
essential role. At 2D-QSI criticality it localizes the pairs and
destroys superfluidity\cite{fisher,herbut} and at the 3D-QSN
transition it destroys superfluidity and the pairs\cite{herbutd}.
In this context we also note that the phase diagram of
La$_{2-x}$Sr$_{x}$CuO$_{4+\delta }$ shown in Fig.\ref{fig4},
reveals that isotope substitution leads to identical behavior,
although less pronounced \cite{zhao214iso}. Since isotope
substitution is accompanied by local lattice distortions, we
conclude that in addition to disorder, lattice degrees of freedom
tune superconductivity in an essential manner. Another essential
facet emerges from the fact that the doping and substitution tuned
flow to the 2D-QSI critical point is associated with an
enhancement of $\gamma _{T}$. Thus, despite the fact that the
fraction $\eta =\left( 1/\lambda _{c}^{2}\left( 0\right) \right)
/\left( 1/\lambda _{a}^{2}\left( 0\right) +1/\lambda
_{b}^{2}\left( 0\right) +1/\lambda _{c}^{2}\left( 0\right) \right)
=1/\left( 1+2\gamma _{T=0}^{2}\right) $\cite {book}, which the
third dimension contributes to the superfluid energy density in
the ground state, is very small, this implies that a finite $T_{c}
$ is inevitably associated with an anisotropic but 3D condensation
mechanism, because $\gamma _{T}$ is finite whenever
superconductivity occurs (see Fig.\ref{fig2}). This points
unambiguously to the conclusion that theories formulated for a
single CuO$_{2}$ plane cannot be the whole story. It does not
imply, however, a 3D pairing mechanism because in the presence of
fluctuations pairing and superfluidity occur separately.

\subsection{Evidence for magnetic field tuned quantum phase transitions}

We have seen (Sec.IIB) that a strong magnetic field destroys
superconductivity at finite temperature. In sufficiently clean
systems this destruction occurs at the first order vortex melting
transition. However, in the presence of disorder, the long-range
order of the vortex lattice is destroyed and the vortex solid
becomes a glass\cite{vortexglass}. The vortex fluid to glass
transition appears to be a second-order transition, signaled by
the vanishing of the zero-frequency resistance in the vortex glass
phase. Since disorder plays an essential role at 2D-QSI and 3D-QSN
criticality, one expects a line $ H_{m}\left( x\right) $ of
vortex-glass to fluid quantum phase transitions, leading to the
schematic phase diagram depicted in Fig.\ref{fig5}.There is the
superconducting phase (S), bounded by the zero-field transition
line, $ T_{c}\left( x,H=0\right) $, the critical lines of \ the
vortex melting or vortex glass \ to vortex fluid transitions,\
$T_{m}\left( x=fixed,H\right) $ and the line of quantum critical
points, $H_{m}\left( x,T=0\right) $. Along this line
superconductivity is suppressed and the critical endpoints
coincide with the 2D-QSI and 3D-QSN critical points at $x_{u}$ and
$x_{o}$, respectively. To fix the critical line $H_{m}\left(
x,T=0\right) $ close to the 2D-QSI and 3D-QSN transitions, we note
that the singular part of the ground state energy density scales
in analogy to Eq.(\ref{eq12}) as
\begin{equation}
f_{s}\left( \delta ,T\right) =\overline{Q}_{D}\left( \overline{\xi
}_{\tau ,0}^{-}\prod\limits_{i=1}^{D}\overline{\xi
}_{i,0}^{-}\right) ^{-1}\delta ^{ \overline{\nu }\left( D+z\right)
}\overline{G}_{D}\left( \overline{z}\right) ,\ \ \
\overline{z}={{\frac{H\overline{\xi }_{a}\overline{\xi }_{b}}{\Phi
_{0}}}}  \label{eq55}
\end{equation}
where $H$ corresponds to a field applied parallel to the $c$-axis.
Supposing that in the ($H,\delta $)-plane there is a continuous
transition first order melting transition. Then the scaling
function $\overline{G}_{D}\left( \overline{z}\right) $ will
exhibit a singularity at some universal value $
\overline{z}=\overline{z}_{m}$ . Thus close the 2D-QSI and 3D-QSN
critical endpoints the quantum critical line (see Fig.\ref{fig5})
is given by,
\begin{equation}
H_{m}\left( \delta \right) =\frac{\overline{z}_{m}\Phi _{0}}{\xi
_{a,0}\xi _{b,0}}\ \delta ^{2\overline{\nu }},  \label{eq60}
\end{equation}
with $\overline{\nu }\approx 1$ (Eq.(\ref{eq39a})) and
$\overline{\nu } \approx 1/2$ (Eq.(\ref{eq39c})) close to 2D-QSI
and 3D-QSN criticality, respectively. Moreover, the singular
behavior of the scaling function $ \overline{G}_{D}\left(
\overline{z}\right) $ must be such as enable the correlation
length when $H\neq 0$ to correspond to the vortex glass
transition, so that
\begin{equation}
\overline{\xi }_{H}=\overline{\xi }_{H,0}\left|
\frac{H-H_{m}\left( \delta \right) }{H_{m}\left( \delta \right)
}\right| ^{-\overline{\nu }_{m}}. \label{eq61}
\end{equation}
\ $\overline{\nu }_{m}$ is the correlation length exponent of the
vortex glass to fluid transition. Noting that $T_{c}\propto \delta
^{\overline{\nu }z}\propto \xi ^{-z}$ (see Eqs.(\ref{eq10}) and
(\ref {eq14}), we obtain with Eq.(\ref{eq61}) for the magnetic
field induced reduction of $T_{c}$ the relation
\begin{equation}
T_{c}\left( H\right) \propto \left( H_{m}\left( \delta \right)
-H\right) ^{ \overline{\nu }_{m}z_{m}}.  \label{eq62}
\end{equation}
$z_{m}$ is the dynamic critical exponent of the vortex glass to
fluid transition.

\begin{figure}[tbp]
\centering
\includegraphics[totalheight=5cm]{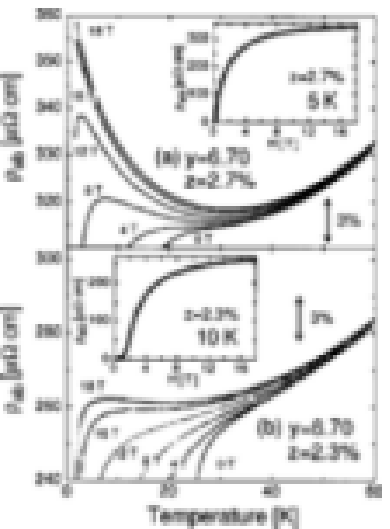}
\includegraphics[totalheight=5cm]{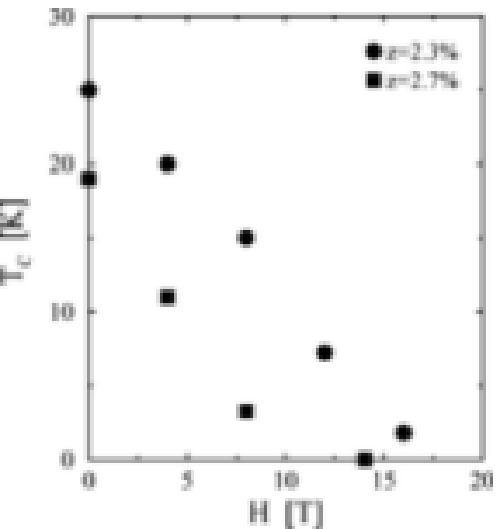}
\caption{Left panel: In-plane resistivity $\protect\rho _{ab}$
versus $T$ at $H=0,\ 4,\ 8,\ 16$ and $18T$ for
YBa$_{2}$Cu$_{3-z}$Zn$_{z}$O$_{y}$. a: $y=6.7,z=2.7\%$ and b:
$y=6.7,z=2.3\%$. Taken from Segawa {\em et al.}
\protect\cite{segawa}. Right panel: Schematic ($H,T$)-phase
diagram of YBa$_{2}$Cu$_{3-z}$Zn$_{z}$O$_{y}$  derived from the
data shown in the left panel.}
\label{fig29}
\end{figure}

The existence of a quantum phase transition line $H_{m}\left(
\delta \right) $ can be anticipated from the in-plane resistivity
data for YBa$_{2}$Cu$_{3-z}$Zn$ _{z}$O$_{y}$ of Segawa {\em et
al.} \cite{segawa} shown in the left panel of Fig.\ref{fig29}.
With increasing magnetic field strength $T_{c}$ is depressed. In
the sample with $y=6.7$ and $z=2.7\%$ the temperature dependence
of $\rho _{ab}$ clearly points to a magnetic field tuned QSI
transition around $ 16<H_{m}\left( y=6.7,z=2.7\%\right) \lesssim
18T$. Since $H_{m}\left( \delta \right) $ scales as $\ \delta
^{2\overline{\nu }}\propto \left( z_{c}-z\right) ^{2\overline{\nu
}}$(Eq.(\ref{eq60})), $H_{m}\left( y=6.7,z\right) $ should
increase with reduced Zn concentration $z$. This behavior is
consistent with the data for $z=2.3\%$, where $H_{m}\left(
y=6.7,z=2.3\%\right) >18T$. The emerging ($H,T$) phase diagram,
consisting of finite temperature transition lines (vortex glass to
fluid transitions) with 2D-QSI critical endpoints is displayed in
the right panel of Fig.\ref{fig29}. In the schematic
($x,H,T$)-phase diagram shown in Fig.\ref{fig5}, these lines
result from cuts near the zero field 2D-QSI-transition with $x$
replaced by $z$.

\begin{figure}[tbp]
\centering
\includegraphics[totalheight=5cm]{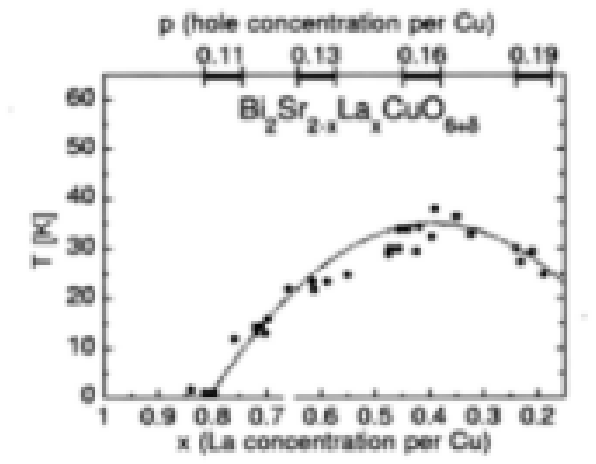}
\includegraphics[totalheight=5cm]{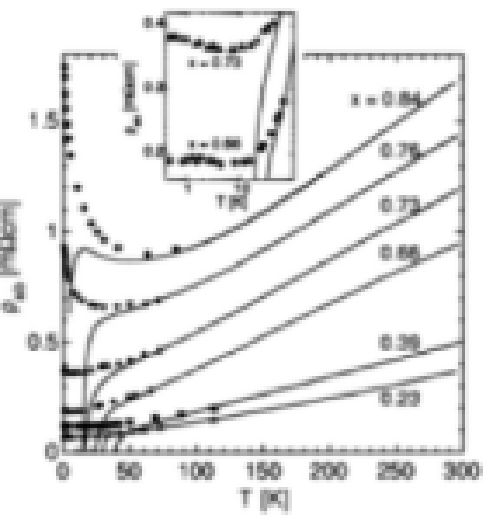}
\caption{Left panel: Phase diagram of
Bi$_{2}$Sr$_{2-x}$La$_{x}$CuO$_{6+\protect\delta }$ in the doping
concentration (x or p)-temperature plane. Taken from Ono {\em et
al.} \protect\cite{ono}. Right panel: Temperature dependence of
$\protect\rho _{ab}$ for a Bi$_{2}$Sr$
_{2-x}$La$_{x}$CuO$_{6+\protect\delta }$ crystal in H = 0 and 60 T
for various dopant concentrations x. The insets shows a clearer
view of the low temperature behavior for x=0.66 and 0.73. Taken
from Ono {\em et al.}\protect\cite{ono}.} \label{fig30}
\end{figure}

\begin{figure}[tbp]
\centering
\includegraphics[totalheight=5cm]{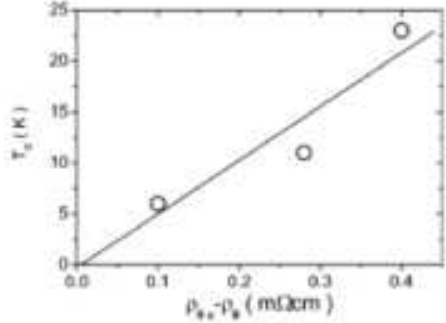}
\caption{$T_{c}$ versus $\protect\rho _{0,c}-\protect\rho _{0}$
for Bi$_{2}$Sr$_{2-x}$La$_{x}$CuO$_{6+\delta }$. Deduced from the
data shown in Fig.\ref{fig30} for x=0.76, 0.73 and 0.66 and $
\protect\rho _{0,c}=0.5$ $m\Omega cm$. The straight line is the
scaling relation (\ref{eq41}) with $z \overline{\protect\nu
}\approx 1$.} \label{fig31}
\end{figure}

\begin{figure}[tbp]
\centering
\includegraphics[totalheight=6cm]{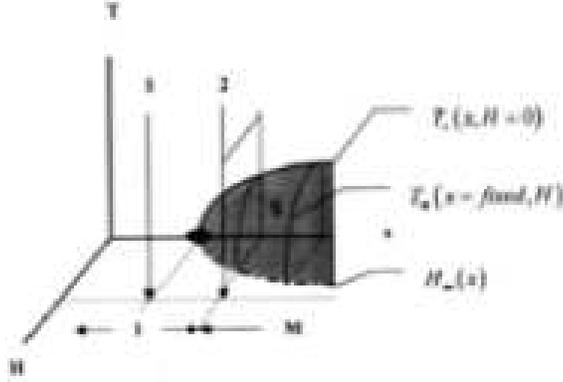}
\caption{Schematic (x,T,H)-phase diagram close to the 2D-QSI
transition. There is the superconducting phase (S), bounded by the
zero-field transition line, $T_{c}\left( x,H=0\right) $, the lines
of the vortex melting or vortex glass to vortex fluid transitions
$T_{m}\left( x=fixed,H\right) $ and the quantum critical line
$H_{m}\left( x\right) $. The critical lines $T_{c}\left(
x,H=0\right) $ and $H_{m}\left( x\right) $ merge in the underdoped
limit ($x=x_{u}$) where a doping driven quantum superconductor to
insulator (QSI) transition ( $\bullet $ ) occurs. The arrows mark
the flows emerging from the experimental data shown in
Fig.\ref{fig31} at $x=0.84$ (1), $x=0.66$ (2).} \label{fig32}
\end{figure}

Related behavior was also observed in La$_{2-x}$Sr$_{x}$CuO$_{4}$
\cite {boebinger} and Bi$_{2}$Sr$_{2-x}$La$_{x}$CuO$_{6+\delta }$
\cite{ono}. In the hole doped La$_{2-x}$Sr$_{x}$CuO$_{4}$
\cite{boebinger} and the electron doped
$\Pr_{2-x}$Ce$_{x}$Cu$_{4+\delta }$ a magnetic field tuned metal
to insulator crossover was observed close to optimum doping, while
in Bi$_{2}$Sr$_{2-x}$La$_{x}$CuO$_{6+\delta }$ \cite{ono} the
crossover sets in well inside the underdoped regime. The phase
diagram of Bi$_{2}$Sr$_{2-x}$La$_{x}$CuO$_{6+\delta }$ is
displayed in the left panel of Fig.\ref{fig30}. In the underdoped
limit ($x\approx 0.84$ ) the transition temperature vanishes and a
2D-QSI transition is expected to occur. Evidence for this
transition emerges from the temperature dependence of $\rho _{ab}$
shown in the right panel of Fig.\ref {fig30}\cite{ono}, taken at
$H=0$ and $60 T$ for six doping levels $x$. To strengthen this
point we plotted in Fig. \ref{fig31} the resulting $T_{c}$ versus
$ \rho _{0,c}-\rho _{0}$ for $x=0.76,0.73$ and $0.66$ and $\rho
_{0,c}\approx 0.5$ $m\Omega cm $ at $H=0$. Apparently there is
suggestive consistency with the 2D-QSI scaling relation
(\ref{eq18}) with $z\overline{\nu }\approx 1$, as well as with the
corresponding data for La$_{2-x}$Sr$_{x}$CuO$_{4}\ $\ and
YBa$_{2}$Cu$_{3}$O$_{7-\delta }$ shown in Fig.\ref{fig16}. As the
$60 T$ data (symbols) are concerned, the samples closest to the
QSI transition ($x=0.76$ and $0.84$) \ exhibit in $\rho _{ab}$ a
pronounced upturn at low temperatures. This points to an
insulating normal state. Indeed, the weak upturn below $6$ $K$ at
$x=0.73$, visible in the inset to Fig.\ref{fig30}, signals the
proximity to the onset of insulating behavior, while the low $T$
behavior of the $x=0.66$ sample with $T_{c}\left( H=0\right) $
$=23 K$ exhibits metallic behavior. Thus, there is a metal to
insulator (MI) crossover. The emerging ($x,T,H$)-phase diagram is
displayed in Fig.\ref{fig32}. The arrows mark the flows emerging
from the experimental data displayed in Fig. \ref{fig30} for
$x=0.84$ (1) and $x=0.66$ (2).

\begin{figure}[tbp]
\centering
\includegraphics[totalheight=5cm]{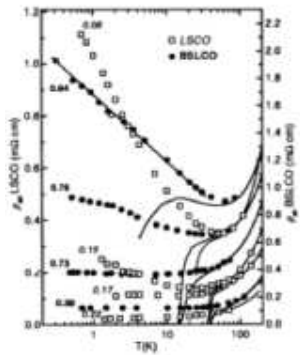}
\caption{Logarithmic plot of $\protect\rho _{ab}\left( T\right) $
for Bi$ _{2} $Sr$_{2-x}$La$_{x}$CuO$_{6+\protect\delta }$ and
La$_{2-x}$Sr$_{x}$CuO$ _{4}$ crystals in H=0T (solid lines) and 60
T (filled circles), labelled by La concentration, $x$. The
straight line indicates consistency with the $ log(1/T)$ behavior
at $x=0.84$. and open circles are $H=30T$ data. La$_{2-x}$
Sr$_{x}$CuO$_{4}$ data in $H=0T$ (dashed lines) and in $60 T$
(open squares), labelled by the Sr concentration x. Taken from Ono
{\em et al.}\protect\cite{ono}.} \label{fig33}
\end{figure}

This differs from the behavior observed in
La$_{2-x}$Sr$_{x}$CuO$_{4}$ where the detectable MI- crossover
appears to set in close to optimum doping. In Fig.\ref{fig33} we
depicted the logarithmic plot of $\rho _{ab}\left( T\right) $ for
Bi$_{2}$Sr$_{2-x}$La$_{x}$CuO$_{6+\delta }$ and La$_{2-x}$Sr$
_{x}$CuO$_{4}$ crystals for $H=0$ and $60 T$ and various doping
concentrations \cite{ono}. Concentrating on the presence or
absence of an upturn, it is clearly seen that in
La$_{2-x}$Sr$_{x}$CuO$_{4}$ the MI-crossover occurs around
$x\approx 0.15$, corresponding to optimum doping. However, it
should be kept in mind that this crossover is nonuniversal. For
this reason, the dopant concentration where insulating behavior is
detectable is expected to be material dependent.

The magneto resistance $\rho _{ab}\left( H,T\right) $ of
underdoped La$_{2-x}$ Sr$_{x}$CuO$_{4}$ films with $x\approx
0.048$ $\left( T_{c}=0.45 K\right) $ and $0.051$ $\left( T_{c}=4
K\right) $ was also studied in magnetic fields up to $0.5 T$ and
at temperatures down to $30mK$ \ \cite{karpinska}. The temperature
dependence of $\rho _{ab}\left( H,T\right) $ of the film closest
to the 2D-QSI transition $\left( x\approx 0.048,\ T_{c}=0.45
K\right) $ was found to be consistent with $\rho _{ab}\propto
exp\left( \left( T/T_{0}\right) ^{1/3}\right) $ below $1 K$ and
for fields from $4$ to $6T$. Thus this study points to
2D-localization. On the other hand, the data of Ando {\em et
al.}\cite{ando} (Fig.\ref{fig33}) appears to be consistent with a
logarithmic divergence of the normal-state resistivity down to
$0.7 K$ for $ x=0.08$ and $H=$ $60 T$, suggesting 2D weak
localization. In any case, the evidence for 2D localization
confirms an essential property of 2D-QSI criticality: disorder
localizes the pairs and destroys superfluidity\cite
{fisher,herbut}.

\section{Thin films and field-effect doping}

There is considerable evidence that the quantum superconductor to
insulator transition in thin films can also be traversed by
changing a parameter such as film thickness, disorder etc.
\cite{book}. As cuprates are concerned, an instructive example are
the measurements of YBa$_{2}$Cu$_{3}$O$_{7-\delta }$ slabs of
thickness $d$ separated by 16 unit cells ($ \approx 187\ A$) of
PrBa$_{2}$Cu$_{3}$O$_{7}$. Due to the large separation the
YBa$_{2}$Cu$_{3}$O$_{7-\delta }$ slabs can be considered to be
essentially to be uncoupled. As shown in Fig.\ref{fig6} in
YBa$_{2}$Cu$_{3}$O$_{7-\delta }$ slabs of thickness $d$ the
transition temperature was found to vary according to
Eq.(\ref{eq1d}). Together with the scaling relation (\ref{eq14})
this points to a 2D-QSI transition with $z\overline{\nu }=1$, in
agreement with our previous estimate (Eq.(\ref{eq39a})). Since
quantum fluctuations alone are not incompatible with
superconductivity, this 2D-QSI transition must be attributed to
the combined effect of disorder and quantum fluctuations. For this
reason it is conceivable that in cleaner films superconductivity
may also occur at and below this value of $d_{s}$, which is close
to the estimate derived from the bulk (Eq.(\ref{eq40})). Recently
this has been achieved by inducing charges by the field-effect
technique, whereby the mobile carrier density can be varied
continuously and reversibly, without introducing additional
sources of disorder\cite {ahn,kawaharo,schon2}. By establishing a
voltage difference between a metallic electrode and the crystal,
charge can be added or removed from the CuO$_{2}$ layers: positive
voltage injects electrons, and a negative voltage injects holes.
This allows to vary both electron and hole charge densities over a
considerable range of interest. Using this technique Sch\"{o}n
{\em et al.}\cite{schon2} converted an insulating thin slab of \
CaCuO$_{2}$ into a superconductor. The success of this technique
relies on the high quality films and on the quality of the
interfaces between the film material and the metallic electrode.
The ($T_{c}$, $x$) phase diagram for CaCuO$_{2}$ is shown
Fig.\ref{fig7}. The doping level was calculated from the
independently measured capacitance of the gate dielectric,
assuming that all the charge is located in a single CuO$_{2}$
layer. The diagram exhibits apparent electron and hole
superconductivity, and bears a strong resemblance to results
observed in other cuprates, in agreement with the empirical
correlation (\ref{eq1c}). Note that the finite temperature
superconductor to normal state transition falls into the 2D-XY
universality class. Thus $T_{c}$ is the Kosterlitz-Thouless
transition temperature. At low doping levels ($ x<0.1$) a
logarithmic divergence of $\rho $ was found. This observation for
n- as well as p-type doping of CaCuO$_{2}$ are in accordance with
a 2D-QSI transition in the underdoped limit, where disorder
localizes the pairs and destroys superfluidity (Eq.(\ref{eq39a})).
Although superconductivity has been observed previously in
SrCuO$_{2}$/BaCuO$_{2}$\cite{norton} and in
BaCuO$_{2}$/CaCuO$_{2}$\cite{balestrino} superlattices with a
maximum $T_{c}$ value of about 80 K, the field effect doping is
not limited by dopant solubility and compositional stability. This
technique allows to study the properties of a given material as a
function of the dopant concentration without introducing
additional disorder and additional defects. Because the charge
carriers appear to be confined in a very thin layer, what's seen
is superconductivity in 2D. Thus, due to the reduced
dimensionality fluctuations will be enhanced over the full range
of dopant concentrations. The phase diagram displayed in
Fig.\ref{fig7} also points to a 2D-QSN transition in the overdoped
limit. A potential candidate is again the disordered metal to
d-wave superconductor transition at weak coupling, considered by
Herbut\cite{herbutd}, with $z=2$ and $\overline{\nu }\approx 1/2$,
and the system is in 2D right at its upper critical dimension.
Clearly, more detailed studies are needed to uncover the
characteristic 2D critical properties in field effect doped
cuprates. Of particular relevance is the confirmation of the
universal linear relationship between $T_{KT}$ and the zero
temperature in-plane penetration depth $\lambda _{ab}\left(
0\right) $ close to the 2D-QSI transition. This relation provides
an upper bound for the attainable transition temperatures.

We saw that in the bulk (see e.g. Fig.\ref{fig2}) and chemically
doped films superconductivity disappears in the 2D-limit, while in
field effect doped CaCuO$_{2}$ it survives. Since chemically doped
materials with different carrier densities also have varying
amounts of disorder, the third dimension is apparently needed to
delocalize the carriers and to mediate superfluidity.

\section{Concluding remarks and comparison with other layered superconductors}

Evidence for power laws and scaling should properly consist of
data that cover several decades in the parameters. The various
power laws which we have exhibited span at best one decade and the
evidence for data collapse exists only over a small range of the
variables. Consequently, though the overall picture of the
different types of data is highly suggestive, it cannot really be
said that it does more than indicate consistency with the scaling
expected near a quantum critical point or a quantum critical line.
Nevertheless, the doping, substitution and magnetic field induced
suppression of $T_{c}$ clearly reveal the existence and the flow
to 2D-QSI and 3D-QSN quantum phase transition points and lines
(see Figs.\ref{fig1}, \ref{fig2}, \ref{fig3} and \ref{fig5}). In
principle, their universal critical properties represent essential
constraints for the microscopic theory and phenomenological
models. As it stands, the experimental data is fully consistent
with a single complex scalar order parameter, a doping induced
dimensional crossover, a doping, substitution or magnetic field
driven suppression of superconductivity, due to the loss of phase
coherence. When the evidence for this scenario persists,
antiferromagnetic and charge fluctuations are irrelevant close to
criticality. Moreover, given the evidence for the flow to 2D-QSI
criticality, the associated 3D-2D crossover, tuned by increasing
anisotropy, implies that a finite $T_{c}$ and superfluid density
in the ground state of bulk cuprates is unalterably linked to a
finite anisotropy. This raises serious doubts that 2D models are
potential candidates to explain superconductivity in bulk
cuprates.Nevertheless, we saw that in sufficiently clean and field
effect doped CaCuO$_{2}$ superconductivity survives in the
2D-limit. Thus, there is convincing evidence that the combined
effect of disorder and quantum fluctuations plays an essential
role and even destroys superconductivity in the 2D limit of
chemically doped cuprates. Thus, as the nature of the quantum
phase transitions in chemically doped cuprates is concerned,
disorder is an essential ingredient. We have seen that the 2D-QSI
transition has a rather wide and experimentally accessible
critical region. For this reason we observed considerable and
consistent evidence that it falls into the same universality class
as the onset of superfluidity in $^{4}$He films in disordered
media, corrected for the long-rangeness of the Coulomb
interaction. The resulting critical exponents, $z=1$ and
$\overline{\nu }\approx 1$ are also consistent with the empirical
relations (Eqs.(\ref{eq1a}), (\ref{eq1b}) and (\ref{eq1c})) and
the observed temperature and magnetic field dependence of the
specific heat coefficient in the limit of zero temperature.  These
properties also point to a 3D-QSN transition with $z=2$ and
$\overline{\nu }\approx 1/2$, describing a d-wave superconductor
to disordered metal transition at weak coupling. Here the disorder
destroys superconductivity, while at the 2D-QSI transition it
localizes the pairs and with that destroys superfluidity. Due to
the existence of the 2D-QSI and 3D-QSN critical points, the
detection of finite temperature 3D-XY critical behavior will be
hampered by the associated crossovers which reduce the temperature
regime where thermal 3D-XY fluctuations dominate. In any case, our
analysis clearly revealed that superconductivity in chemically
doped cuprate superconductors is a genuine 3D phenomenon and that
the interplay of disorder (anisotropy) and superconductivity
destroys the latter in the 2D limit. As a consequence, the
universality of the empirical correlations reflects the flow to
2D-QSI and 3D-QSN criticality, tuned by chemical doping,
substitution and an applied magnetic field. A detailed account of
the flow from 2D-QSI to 3D-QSN criticality is a challenge for
microscopic theories attempting to solve the puzzle of
superconductivity in these materials. Although the mechanism for
superconductivity in cuprates is not yet clear, essential
constraints emerge from the existence of the quantum critical
endpoints and lines. However, much experimental work remains to be
done to fix the universality class of the 2D-QSI and particularly
of the 3D-QSN critical points unambiguously.

\begin{figure}[tbp]
\centering
\includegraphics[totalheight=9cm]{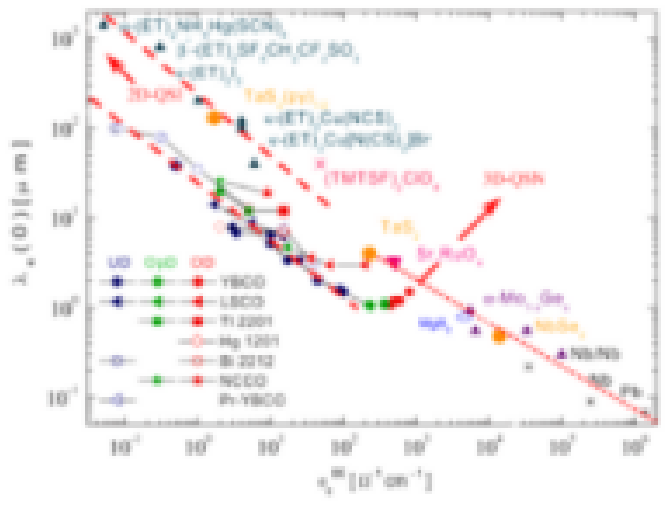}
\caption{$\protect\lambda _{c}\left( 0\right)$ versus
$\protect\sigma _{c}^{DC}$ for a variety of superconductors
partially collected by Dordevic{\em et al.}\protect\cite
{dordevic}. YBa$_{2}$Cu$_{3}$O$_{7-\protect\delta}$
(YBCO)\protect\cite{ybco1,ybco2,ybco3,ybco4,ybco5},
La$_{2-x}$Sr$_{x}$CuO$_{4+\protect\delta}$
(LSCO)\protect\cite{ybco3,lsco6,lsco7},
HgBa$_{2}$CuO$_{4+\protect\delta }$ (Hg1201)\protect\cite{hg},
Tl$_{2}$Ba$_{2}$CuO$_{6+\protect\delta}$
(Tl2201)\protect\cite{tl9,tl10,tl11},
Bi$_{2}$Sr$_{2}$CaCu$_{2}$O$_{8+\protect\delta}$
(Bi2212)\protect\cite{bi12,bi13},
Nd$_{2-x}$Ce$_{x}$CuO$_{4+\protect\delta}$(NCCO)\protect\cite{nd14}.
Blue points -underdoped (UD), green -optimally doped (OpD) and red
-overdoped (OD). Transition metal
dichalcogenides\protect\cite{tmt1,tmt2}; (ET)$_{2}$X
compounds\protect\cite{su1,carrington,shibauchi2,dressel,taniguchi1,wanka,kajita,su2,prozorov};
(TMTSF)$_{2}$ClO$_{4}$\protect\cite{garoche,le,finley,trey,onabe,kennedy,thompson};
Sr$_{2}$RuO$_{4}$ \protect\cite{maeno,yoshida};
MgB$_{2}$\protect\cite{mgb,angst,finnemore};
Nb\protect\cite{nb19,nb20}; Pb\protect\cite{nb20}; Nb Josephson
junctions \protect\cite{nbjos};
$\protect\alpha$Mo$_{1-x}$Ge$_{x}$\protect\cite{moge}. The dotted
line is Eq.(\ref{eq65}) and the dashed ones Eq.(\ref{eq32}). The
red arrows indicate the flow to 2D-QSI and for overdoped
YBa$_{2}$Cu$_{3}$O$_{7-\delta}$ to 3D-QSN criticality,
respectively.} \label{fig34}
\end{figure}

In conclusion we note that universal properties emerging from
thermal and quantum fluctuations are not restricted to cuprate
superconductors. Potential candidates are the highly anisotropic
organic superconductors which are close to a metal-insulator
boundary. A glance to Fig.\ref{fig34} shows that the correlation
between $\lambda _{c}\left( 0\right) $ and $ \sigma
_{c}^{DC}\left( T_{c}^{+}\right) $ is particularly rewarding. In
various classes of layered superconductors, including organics,
transition metal dichalcogenides and cuprates, $\lambda _{c}\left(
0\right) $ is systematically suppressed with increase of normal
state conductivity $\sigma _{c}^{DC}\left( T_{c}^{+}\right) $.
Without thermal and quantum fluctuations the DC conductivity is
given by
\begin{equation}
\sigma _{c}^{DC}=\frac{ne^{2}}{m_{c}}\tau _{tr}=\frac{c^{2}\tau
_{tr}}{4\pi \lambda _{c}^{2}\left( 0\right) }.  \label{eq64}
\end{equation}
$\tau _{tr}$ is the mean scattering time of the normal electrons
in transport properties with number density per unit volume $n$
and effective mass $m_{c}$. Provided that the variations of $\tau
_{tr}$ are small, superconductors where fluctuations can be
neglected (mean-field superconductors) are then characterized by
the dotted line in Fig.\ref{fig34}, which is
\begin{equation}
\lambda _{c}\left( 0\right) \propto \left( \sigma _{c}^{DC}\right)
^{-1/2}. \label{eq65}
\end{equation}
It agrees with the data of mean-field superconductors rather well.
Interestingly enough, MgB$_{2}$ and Sr$_{2}$%
RuO$_{4}$ appear to fall into this class, as well. Prominent and
systematic deviations from this behavior occur for the
highly anisotropic organics and underdoped cuprates. As $%
\lambda _{c}\left( 0\right) $ increases with the reduction of the
normal state conductivity $\sigma _{c}^{DC}$, we observe that the
data tends to scatter around two branches, consistent with
$\lambda _{c}\left( 0\right) =\Omega _{s}\left( \sigma
_{c}^{DC}\right) ^{-3/4}$, which is the scaling relation
(\ref{eq32}) with $z=1$, describing the critical behavior close to
2D-QSI critical points. The most prominent deviations from these
trends occur for overdoped cuprates (see also Fig.\ref{fig24}). In
particular the data for overdoped YBa$_{2}$Cu$_{3}$O$_{7-\delta
\text{ }}$(YBCO) exhibit an upturn, signaling the crossover to
3D-QSN criticality (Eq.(\ref{eq37})). Since organics undergo
superconductor to insulator transitions\cite {nakazawa}, are
highly anisotropic and $\gamma $ can be varied over a rather
extended interval, i.e. $\gamma _{T_{c}}=180$ for $\kappa
$-(ET)$_{2}$ Cu[N(CS)$_{2}$ ]Br, $\gamma _{T_{c}}=350$ for $\kappa
$-(ET)$_{2}$ Cu(NCS)$_{2}$\cite{kawamata} and. $\gamma _{T_{c}}=2\
10^{3}$ for $\alpha $- (BEDT-TTF)$_{2}$NH$_{4}$Hg(SCN)$_{4}$, it
becomes clear that the organics line in Fig.\ref{fig34} indicates
the flow of this class of superconductors to 2D-QSI criticality,
where the scaling relation (\ref{eq32}) applies.

Thus unlike the organics the cuprates undergo a doping tuned
crossover from 2D-QSI to 3D-QSN criticality, where $\sigma
_{c}^{DC}\left( T_{c}^{+}\right) $ and $\lambda _{c}\left(
0\right) $ tend to infinity, while $T_{c}$ vanishes. These
critical points are attained in the underdoped and overdoped
limit, respectively, where $T_{c}$ vanishes. In this crossover the
cuprates either pass the organics or the dotted mean-field line,
spanned by conventional superconductors (see Fig.\ref{fig34}).
YBa$_{2}$Cu$_{3}$O$_{7-\delta }$(YBCO) and
La$_{2-x}$Sr$_{x}$CuO$_{4+\delta }$(LSCO) cross the dotted
mean-field line in the neighborhood of TaS$_{2}$,
Sr$_{2}$RuO$_{4}$ and MgB$_{2\text{ }}$. In this region of the
$\lambda _{c}\left( 0\right) $-$\sigma _{c}^{DC}\left(
T_{c}^{+}\right) $ plane fluctuations do not play a dominant role.
Indeed, Sr$_{2}$RuO$_{4}$ exhibits Fermi liquid normal state
properties\cite{bergemann}. However, overdoped cuprates flow to
3D-QSN criticality, where disorder destroys the phase coherence
and the pairs. On the contrary, the organics evolve rather
smoothly from the mean-field regime, to which Pb, Nb,
Nb-junctions, $\alpha $-Mo$_{1-x}$Ge$_{x} $ and the
dichalcogenides belong. Thus, these flows provide a perspective of
the regimes where fluctuations and disorder are essential.

It is a pleasure to thank H.Keller and K.A. M\"{u}ller for
numerous helpful
discussions.

\end{document}